# Beyond Falsehoods 2026

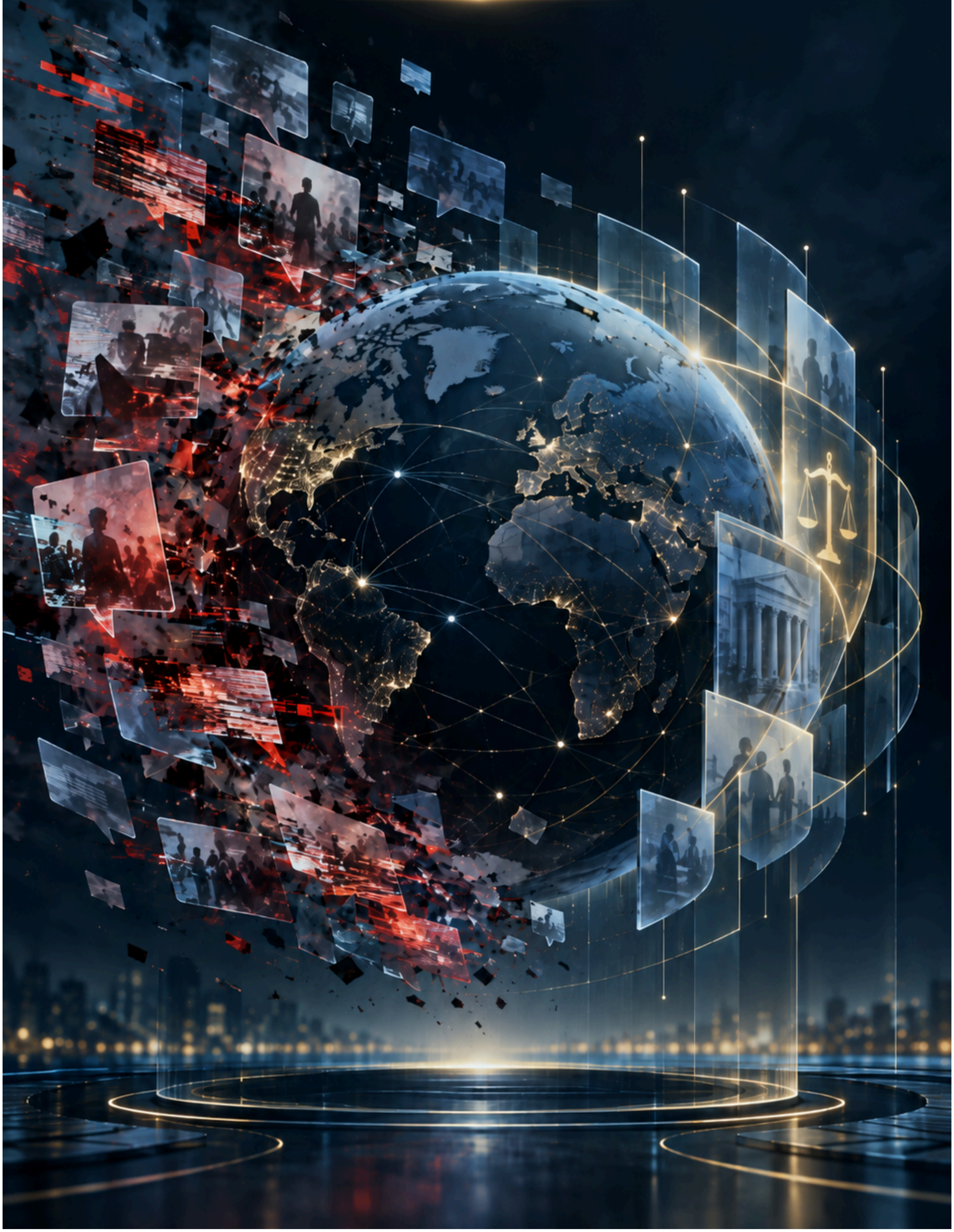

# Beyond Falsehoods 2026

## Global Trends in the Regulation of Misinformation and Disinformation

**Fakhar Abbas, Simon Chesterman, Eka Nugraha Putra**

## Executive Summary

- This report examines **global legislative and regulatory responses to misinformation, disinformation, and mal-information** (MDM) between 1995 and 2025. Drawing on an expanded and updated dataset of 409 coded regulatory and policy entries, including hard and soft law, policy measures, and proposed or draft instruments, it analyses their growth, geographic distribution, substantive focuses, and regulatory designs, which represent the various approaches taken against MDM, as well as early evidence of their implementation.

- The central finding is **sustained and accelerating regulatory growth**. Annual activity increased from 11 instruments in 2016 to 46 in 2023, 67 in 2024, and 71 in 2025. The number of instruments currently in force has reached 282, despite 34 measures having been repealed, withdrawn, or judicially invalidated. This growth has also been accompanied by a shift in regulatory focus. Earlier measures relied heavily on broad prohibitions of "false information," whereas more recent measures i**ncreasingly target specific harms**, including deepfakes, election interference, non-consensual intimate imagery, and child safety. Regulatory **responsibility is also increasingly being placed on platforms** rather than solely on individual creators and users.

- The report classifies these responses into **six regulatory models**: truth-policing, criminal harm prevention, platform systems governance, victim-centred rights and remedies, provenance regulation, and resilience-based approaches. Criminal harm prevention and truth-policing remain among the most common. Since 2023, however, new regulatory activity has increasingly concentrated on platform systems governance, victim-centred remedies, and provenance regulation. This indicates a broader shift **from assessing the truthfulness** of individual claims **towards regulating the systems**, technologies, and actors through which information is created and distributed.

- The report also moves beyond the formal content of legislation to examine **how regulation operates in practice**. Illustrative cases, such as Singapore's POFMA, the European Union's Digital Services Act, South Korea's deepfake-crime amendments, and the judicial invalidation of truth-determination measures in India and Mongolia, show that enforcement outcomes, unintended consequences, and institutional capacity can differ substantially from what legislation provides on paper. The evidence suggests that regulatory effectiveness depends not only on the scope of formal legal powers, but also on the capacity of institutions to implement and enforce them consistently.

- The report concludes by identifying priorities for the future. These include converting unstructured information on informational harms, safeguards and institutional arrangements into comparable variables; systematically recording implementation, enforcement, and judicial outcomes alongside formal legal powers; extending the analysis of factors associated with regulatory design and implementation; expanding the use of non-English-language sources; and maintaining the dataset through annual updates, with substantive updates planned for mid-2027 and mid-2028.

**About the report:** This report is part of the Digital Information Resilience: Restoring Trust and Nudging Behaviours in Digitalization Programme at the National University of Singapore, known informally as Information Gyroscope or iGyro. The research is supported by the Ministry of Education, Singapore, under its MOE AcRF TIER 3 Grant (MOE-MOET32022-0001).

The regulatory dataset covers instruments through 31 December 2025; the discussion of implementation and subsequent developments is current to 31 August 2026 unless otherwise specified.

**About the authors:** Fakhar Abbas is Senior Research Fellow at the Centre for Trusted Internet and Community, National University of Singapore; Simon Chesterman is David Marshall Professor of Law at the National University of Singapore; Eka Nugraha Putra is Research Fellow at the Centre for Trusted Internet and Community, National University of Singapore.

The report benefited from research assistance from Jiang Yu Hang, Shireen Lee, Oh Jong Wee, Elizabeth Ong, Clarie Sng, and Sripratak Thanakorn.

# Contents

# 1. Information Disorder as a Regulatory Problem

## 1.1. Misinformation, Disinformation, Mal-information and Adjacent Informational Harms

The impact of widely shared harmful information has been recognized as one of the top risks facing the planet,[1] on a level comparable to other global crises such as climate change and armed conflict. The problem of false information is not new, but it has been augmented by technology. While the internet enabled the widespread dissemination of false information, Artificial Intelligence (AI) now massively amplifies its spread by making content creation easier, faster, and more targeted.

Concerns about the threat have led to discussions about regulation. However, regulatory responses have been fragmented across countries, with many jurisdictions addressing false information within broader offences such as defamation and hate speech. Moreover, although AI-generated content has dramatically increased the volume of harmful online information, these interventions continue to focus on adjacent harms, such as defamation or incitement, rather than directly addressing the harms caused by information disorder itself.

Although the terms "misinformation" and "disinformation" gained prominence through the framework established by Hossein Derakhshan and Claire Wardle in 2017,[2] the underlying concepts have a longer history. In this report, misinformation is defined as false or misleading information shared without malicious intent, such as urban legends or rumors. Disinformation is false or misleading information deliberately created or disseminated to deceive, manipulate, or advance a particular agenda. Malinformation is genuine information weaponized to cause personal harm through prejudice, hate speech, or harassment. While discussions of information disorder traditionally focus on misinformation and disinformation, malinformation represents a critical third category. Together, these three types constitute MDM (misinformation, disinformation, and malinformation). Table 1 presents the framework for categorizing these forms of harmful information.

**Categorization of Potentially Harmful Information**

| | | Veracity: *True* | Veracity: *False* |
|---|---|---|---|
| **Intent** | *Benevolent/neutral* | — | Misinformation |
| **Intent** | *Malevolent* | Mal-information | Disinformation |

**Table 1:** Categorization of Potentially Harmful Information.

[1] World Economic Forum. (2026). *The Global Risks Report 2026 21st Edition: Insight Report*. Available at https://reports.weforum.org/docs/WEF_Global_Risks_Report_2026.pdf.

[2] Wardle, C., & Derakhshan, H. (2017). *Information disorder: Toward an interdisciplinary framework for research and policymaking* (Report No. DGI(2017)09). Council of Europe. Available at https://edoc.coe.int/en/media/7495-information-disorder-toward-an-interdisciplinary-framework-for-research-and-policy-making.html

The differences in these categories are significant. The distinction in intent helps identify the harm caused by MDM. Therefore, regulation of MDM should go beyond determining whether information is true or false and consider whether actual harm occurs when such information is disseminated online.

## 1.2. Focus on Harms and Regulatory Choices

Several issues arise in understanding how countries respond to MDM. These include whether countries amend existing laws or enact new legislation to regulate MDM; whether they use hard law (such as statutes or acts) or soft law (such as policies or guidelines); and whether they pass any legislation at all. These varied legislative responses are significant for assessing whether addressing MDM is a primary concern for states when regulating harmful information and the technologies that enable its creation and widespread dissemination. Such assessment considers whether laws include sanctions and stringent provisions, or whether countries instead adopt a more flexible approach through soft law and frequent revisions in response to rapidly evolving technology.

Given these different approaches, this report examines how MDM-related harms are currently addressed within broader categories of informational harm, particularly focusing on what legislation targets. Key considerations include whether legislation uses the specific terms "misinformation" and "disinformation," or whether it instead regulates protected interests that MDM may threaten, such as national security, election integrity, public health, protection of minors, or defense against hostile information campaigns.

## 1.3. Tensions between Safety, Institutional Integrity, Free Expression and Political Abuse

The discourse on protected and unprotected expression has been central to debates over regulating MDM. Many jurisdictions have laws on defamation, hate speech, and hate crime. Although such laws restrict expression to protect interests such as reputation and public order, their implementation is simultaneously undergoing intense public and legal scrutiny. A difficult challenge in regulating online expression is determining how harmful online expression may affect not only individuals and communities, but also institutions and even states.

The various harms identified and regulated in the legislation reveal a dilemma that countries face in protecting society from the harms caused by MDM. If the laws passed are too strict, they can also be detrimental to human rights, such as freedom of expression. At the same time, the state needs to educate and equip citizens with adequate digital literacy skills to verify and critically consume information, while also protecting state institutions and the media as reliable sources of information. This is especially important given the potential for false information to be abused in ways that influence people.

Furthermore, laws and policies that are drafted and passed in response to MDM should ideally not only impose sanctions as a tool to control and reduce harmful digital information, but also empower citizens in combatting MDM.

This report summarizes the regulatory approaches that have been taken by countries around the world. It is part of "Digital Information Resilience: Restoring Trust and Nudging Behaviours in Digitalization"[3], an interdisciplinary programme that addresses the production, dissemination and consumption of MDM. Regulation and policy aspects in this project play an important role in how these affect the supply and demand of digital information, particularly how to prevent the large amount of creation and dissemination of harmful information.

### 1.4. Expanded Scope

This report draws on a global legislative database,[4] presented here as a regulatory landscape dataset of laws and policies addressing online information. The dataset focuses on instruments passed since 1995, which marked the beginning of significant efforts to regulate an information space that was increasingly online.

Subsequent years saw significant increases in the legislation governing MDM. Key moments coincided with the Cambridge Analytica Scandal of 2016-2018 and the COVID-19 pandemic of 2020. The release of ChatGPT by OpenAI in 2022 also coincided with a further increase in the number of laws addressing MDM.

Expanding on earlier research,[5] this report furthers the discussion and analysis of trends in legislative responses, based on an updated database. The report discusses not only enacted hard law, but also drafted and repealed laws and soft-law instruments. The inclusion of these various interventions aims to discuss the focus of the state in regulating harms caused by MDM and which approaches have been taken by these instruments.

## 2. Methodology and the Regulatory Landscape Dataset

This report draws on a regulatory landscape dataset that systematically identified and coded 409 regulatory and policy instruments addressing online MDM. As the report compares regulatory developments across jurisdictions and over time, a clear and consistent methodology is necessary to ensure that the findings are reliable, comparable, and reproducible.

This section explains the dataset's scope, selection criteria, units of analysis, coding framework, data sources, quality checks, and analytical methods. It also outlines the dataset's main limitations and how they are addressed. Finally, it distinguishes between descriptive findings drawn directly from the coded data and statistical analyses examining relationships between variables. Together, these methodological foundations support the empirical analysis in the sections that follow.

### 2.1. Inclusion and Exclusion Rules

Determining whether an instrument addresses MDM requires clear criteria that can be applied consistently across jurisdictions. Some instruments regulate MDM indirectly through broader areas such as public order, national security, or criminal law. Others directly target MDM but use terms such as "fake news," "false information," "misleading content," or "hostile information campaigns."

[3] Centre for Trusted Internet and Community. (n.d). *iGyro: Information Gyroscope*. Available at https://ctic.nus.edu.sg/igyro/.

[4] Centre for Trusted Internet and Community. (n.d). Global Legislation on Misinformation, Disinformation, and 'Fake News'. Available at https://ctic.nus.edu.sg/igyro/demo-legislative-map.html.

[5] Chesterman, S. (2024). Lawful but Awful: Evolving Legislative Responses to Address Online Misinformation, Disinformation, and Mal-Information in the Age of Generative AI. *The American Journal of Comparative Law*, 72 (4), 933-965.

The dataset therefore classifies instruments by their regulatory purpose rather than the terminology they use. It includes instruments governing the creation, dissemination, or regulation of false or misleading online content. Three refinements were introduced for this report.

- Instruments are included if they specifically regulate the online spread of false, misleading, or harmful information. Statutes, such as criminal code or penal code provisions, are included only when explicitly applied to online expressions.
- Laws concerning obscenity, incitement, hate speech, defamation, or national security are excluded unless they contain specific provisions addressing MDM.
- The dataset covers more than enacted national legislation. It also includes draft and proposed legislation, subsidiary regulations, and non-binding instruments, such as codes of practice, guidelines, declarations, action plans, and selected policy reports. This broader scope captures the regulatory lifecycle from proposal and implementation to amendment, repeal, or withdrawal.

These inclusion criteria also explain differences in the reported totals. When the analysis is restricted to binding measures—such as enacted legislation, decrees, and regulations—the dataset records 205 instruments by the end of 2023. By the end of 2025, the number had increased to 271, spanning a broader range of jurisdictions.

## 2.2. Units of Analysis

The primary unit of analysis is the individual regulatory or policy instrument. Each law, bill, regulation, code, declaration, or policy measure is coded as a separate observation for the jurisdiction in which it applies. Amendments are recorded separately and linked to the original instrument. This approach distinguishes the initial adoption of a regulatory framework from its later development, allowing the dataset to track both the spread of regulatory approaches across jurisdictions and their evolution over time. Each instrument is coded using more than 70 variables organised into five categories:

- Jurisdictional and structural information, including the jurisdiction, UN regional grouping, instrument type, year proposed or first adopted, year it entered into force, and year repealed, where applicable.
- Substantive coverage, spanning issues such as misinformation, disinformation, hostile information campaigns or foreign interference, terrorism or national security, protection of minors, elections, public health or COVID-19, together with a free-text field capturing emerging topics such as deepfakes, intimate-image abuse, fraud, and impersonation.
- Regulatory targets, including content creators, online platforms, media organizations, individuals who knowingly share false information, and individuals who share it unintentionally or carelessly.
- Regulatory powers and safeguards, including powers to determine whether information is true or false, require corrections or content removal, impose fines, or order imprisonment. The dataset also records the bodies exercising these powers, the availability of judicial appeals, and details of exemptions, oversight mechanisms, and other procedural safeguards.

- Country-level contextual indicators, including GDP per capita, internet and mobile penetration, Freedom House Global Freedom scores for 2023 and 2024, and the World Bank Worldwide Governance Indicators for government effectiveness, regulatory quality, rule of law, and political stability. These indicators provide additional context for comparing regulatory approaches across jurisdictions.

## 2.3. National, Subnational, Regional and International Instruments

The analysis focuses primarily on national regulatory instruments, as they provide the most consistent basis for comparison across jurisdictions. The dataset covers 201 entities: the 193 UN Member States, several territories, and one regional body included for reference. Of these, 170 have at least one coded instrument, while 151 have at least one binding legal instrument.

Subnational measures are excluded because systematic and comparable coverage across all jurisdictions is not currently feasible. This limitation is particularly relevant to federal systems such as Australia, India, Nigeria, and the United States, where states or provinces may play an important role in regulating information disorder. The findings therefore describe national regulatory frameworks rather than the full extent of regulatory activity within each jurisdiction.

Regional and international instruments, including the European Union's Digital Services Act and AI Act, as well as measures adopted by the Council of Europe, ASEAN, and the African Union, are retained in the dataset but excluded from national-level counts. Their inclusion captures the broader regulatory context and supports analysis of how regional and international frameworks shape national responses.

## 2.4. Hard Law, Proposals, Soft Law and Policy

Governments address information disorder through legal and policy instruments that vary in status and function. Distinguishing among these instruments helps clarify both the scale of regulatory activity and the strategies used to govern online information. For consistent comparison across jurisdictions and over time, each instrument is assigned to one of five categories based on the free-text "type of document" field and standardised coding rules.

| Category | What it captures | n (all years, to 2025) | % of coded instruments |
|---|---|---|---|
| Hard law / binding instrument | Enacted laws, decrees, regulations, subsidiary legislation, organic laws, directives, penal-code reforms | 271 | 67% |
| Soft law/policy | Codes of practice/ethics, guidelines, declarations, action plans, frameworks, resolutions, conventions | 56 | 14% |
| Proposed / draft | Bills, draft laws, draft bills and formal proposals not yet enacted | 45 | 11% |
| Commentary/report | Policy reports, briefs, factsheets and equivalent non-binding analytical outputs | 32 | 8% |
| Other/unclear | Entries not cleanly classifiable from the source description | 3 | 1% |

**Table 2:** Composition of the coded dataset by instrument type
(n = 407 coded instruments, all years; percentages rounded).

This classification distinguishes the growth of binding regulation from the wider use of non-binding policy instruments. It also captures different stages and forms of regulatory development. Hard law instruments impose legally enforceable obligations, while proposed and draft measures indicate areas of active legislative development. Soft law and policy instruments rely instead on coordination, guidance, voluntary compliance, or institutional capacity-building. Commentary and reports provide analysis that may inform later policy or legislative action.

Examining these categories together shows how regulatory responses may develop from policy discussion and legislative proposals to enactment, implementation, amendment, and, where applicable, repeal or withdrawal. The proposal pipeline offers a further indication of possible future developments. Of the 45 proposed or draft instruments in the dataset, seven were subsequently enacted and 34 remain under legislative consideration. Although their outcomes cannot be predicted, pending proposals provide an early indication of emerging regulatory priorities. This classification underpins the analysis of regulatory growth and diffusion in Section 3 and the changing regulatory models examined in Section 7.

### 2.5. Sources, Language Limitations and Verification

The dataset draws on multiple sources to improve coverage, accuracy, and comparability across jurisdictions. It builds on established datasets of regulatory responses to misinformation and disinformation and was expanded through systematic searches of each jurisdiction using a consistent search strategy. The 2024–2025 update also drew on official legislative databases, government gazettes, and specialist sources covering recent developments, including implementation of the European Union's Digital Services Act and AI Act, the United Kingdom's Ofcom codes, and emerging legislation on deepfakes and other AI-generated content.

Most searches were conducted in English, with non-English legislative materials translated where necessary. Coverage may therefore be less complete in jurisdictions where relevant materials are published primarily in other languages or are not readily accessible online. When no relevant instrument was identified, the jurisdiction was recorded as "no data to date." This classification indicates a lack of verified data rather than confirming that no relevant instrument exists. To support transparency and reproducibility, each entry includes a source URL and, where available, a brief description based on the primary source. For quality assurance, a second expert independently cross-checked a sample of entries from each UN regional group.

### 2.6. Coding Reliability and Treatment of Amendments and Repeals

Because regulatory frameworks change over time, the dataset uses consistent rules for recording amendments, repeals, withdrawals, and judicial invalidation. These rules are important for distinguishing new regulatory measures from changes to existing ones and for interpreting the trend analyses presented in Sections 3 and 8.

- Treatment of amendments. Substantive amendments are coded as separate instruments, linked to the original entries, and assigned their own year of adoption. This preserves the sequence of regulatory developments. Amendments involving only technical or administrative changes, such as renumbering provisions without changing their substantive effect, are excluded.

- Treatment of repeals and invalidated measures. Repeals, withdrawals, superseded instruments, and successful constitutional or judicial challenges are recorded in a dedicated status field, together with explanatory notes where publicly available. At the time of writing, the dataset includes 34 instruments classified as repealed, withdrawn, superseded, or judicially invalidated.

To improve coding reliability, the research team independently coded and reviewed entries using a shared framework. Disagreements were resolved through discussion and verification against the original legal or policy sources, helping ensure consistent coding across jurisdictions and instrument types.

### 2.7. What is Descriptive versus Inferential

The report distinguishes between descriptive and inferential analysis to avoid drawing conclusions beyond what the data support. Descriptive analysis identifies patterns within the dataset, whereas inferential analysis examines relationships between variables.

- Descriptive analysis reports frequencies, proportions, trends over time, and cross-tabulations. These findings show how the coded instruments are distributed and how regulatory approaches have evolved, subject to the dataset's coverage and coding limitations.
- Inferential analysis uses statistical techniques, including correlation and group comparisons, to examine associations between variables. These relationships may identify patterns requiring further investigation but do not establish causation.

The report therefore makes no causal claims. Any possible explanation for an observed relationship is presented as a hypothesis for future research rather than a confirmed causal conclusion. The next section applies this methodology to examine regulatory growth, geographic distribution, and regulatory phases.

## 3. Growth, Geography and Regulatory Phases

Global responses to MDM have expanded substantially over the past three decades. This growth is evident not only in the number of instruments adopted but also in the wider range of legal and policy approaches used to govern the online information environment. Examining these developments across jurisdictions and over time shows how regulatory responses have changed alongside technological, political, and social conditions.

This section traces regulatory developments through the end of 2025. It first examines cumulative and annual growth, then distinguishes initial adoption from later amendments to separate the spread of regulation across jurisdictions from the development of existing frameworks. It subsequently analyses regional patterns and the main phases of regulatory change before comparing enacted and proposed measures to identify emerging trends.

### 3.1. Global Cumulative and Annual Trends

The dataset shows a marked acceleration in global regulatory activity. The annual number of instruments adopted or proposed reached 46 in 2023, 67 in 2024, and 71 in 2025—the three highest totals in the dataset. Annual activity in 2025 was approximately 11 times the 2013-2017 average of about six instruments per year. Over the same period, the cumulative number of binding legal instruments increased from 205 at the end of 2023 to 271 at the end of 2025, representing a 32% increase. The dataset also records 34 instruments that were repealed, withdrawn, superseded, or invalidated by courts.

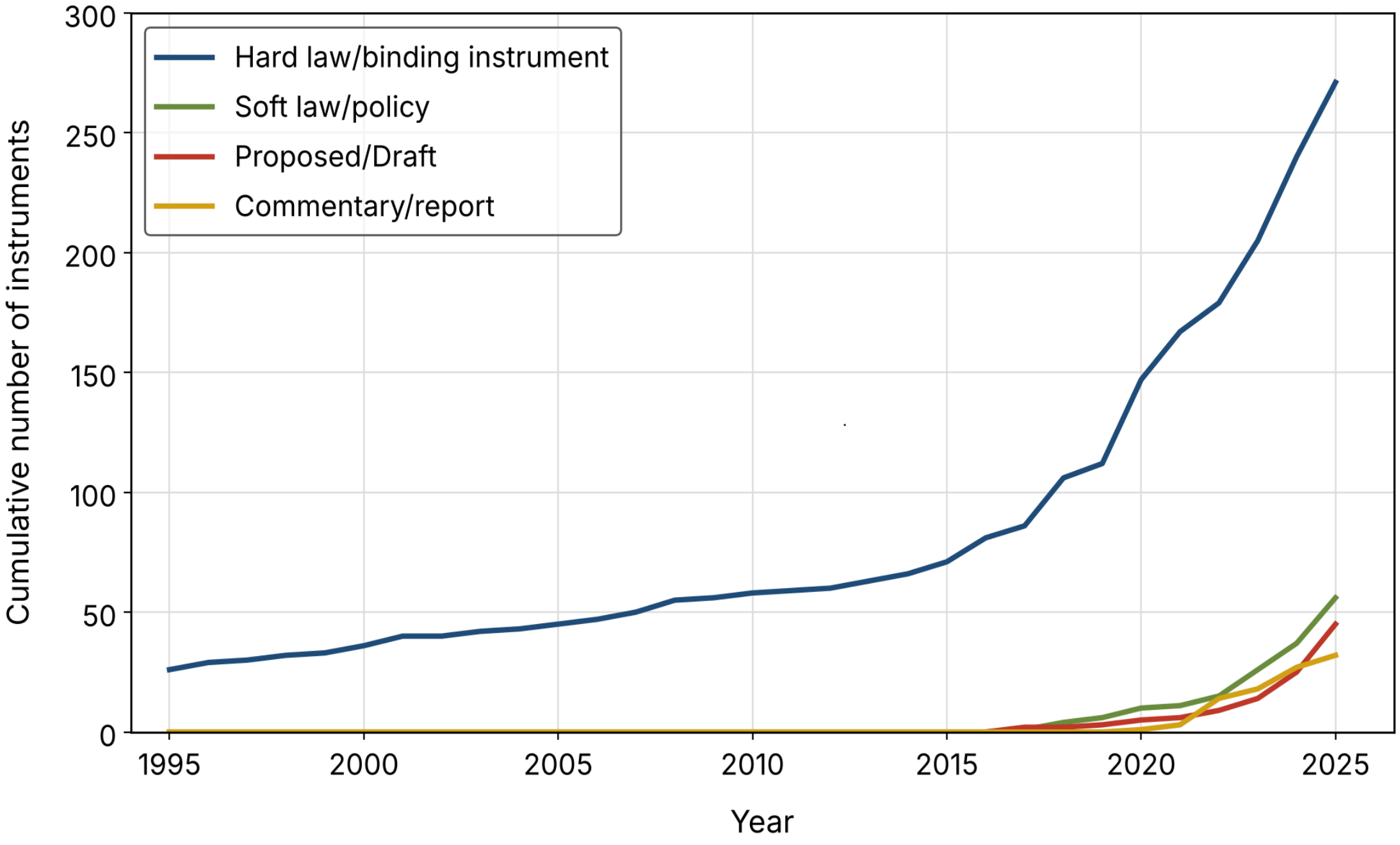

**Figure 1:** Cumulative growth of coded instruments, 1995–2025, by category (hard law, soft law/policy, proposed/draft, commentary/report).

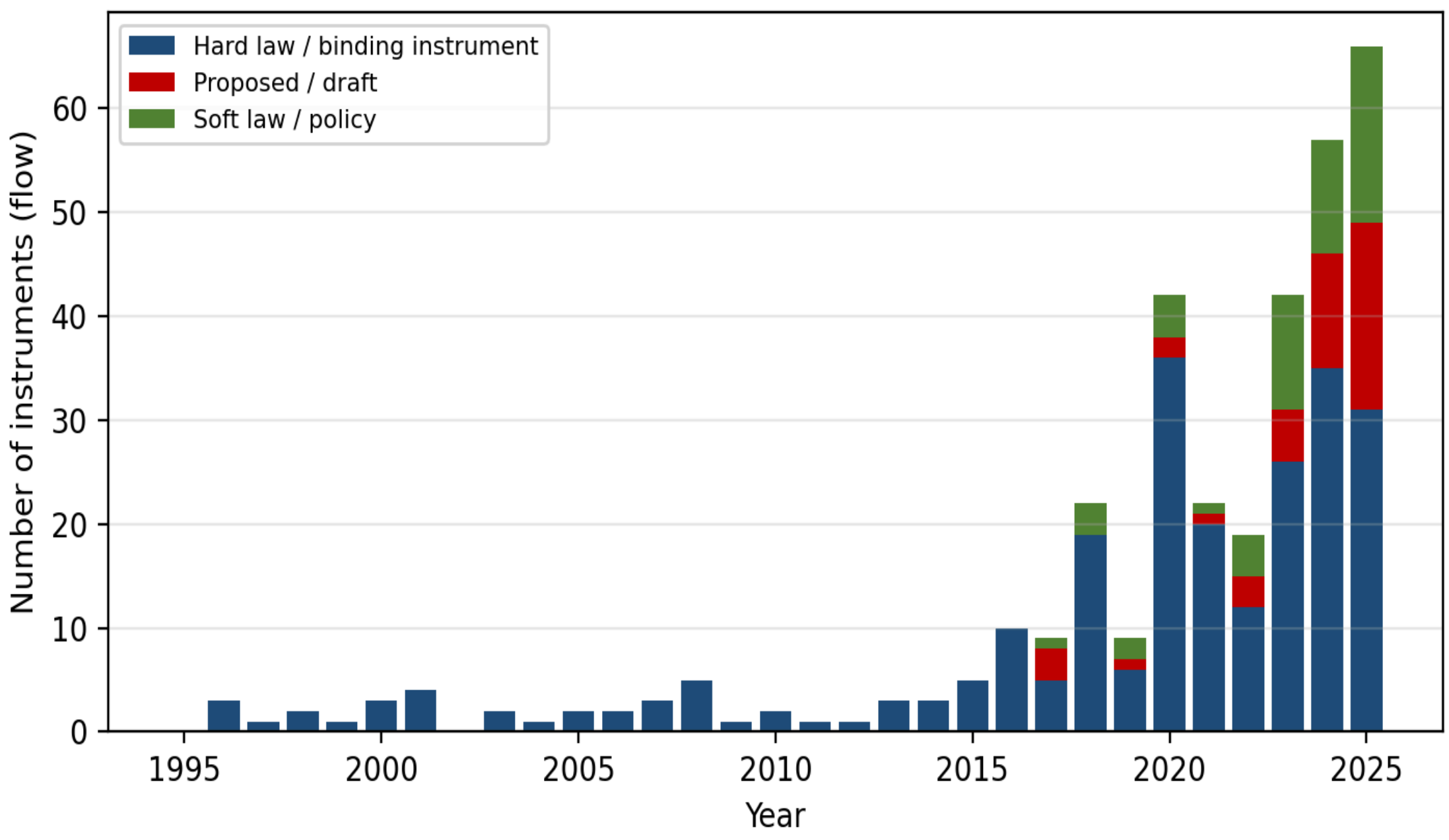

**Figure 2:** New instruments adopted or proposed per year, 1995–2025, decomposed by category.

Two findings stand out. First, binding legal instruments continue to represent the largest share of regulatory activity. Since 2020, however, their relative share has declined as governments have increasingly adopted soft-law measures and introduced legislative proposals. Regulatory responses are therefore becoming more varied, with non-binding measures increasingly supplementing formal legislation.

Second, improved data coverage alone does not explain the recent growth. The dataset records 92 binding legal instruments, adopted between 2023 and 2025,—roughly twice the 45 adopted throughout 1995–2015. The scale and pace of recent activity therefore indicate a substantial expansion of government efforts to address information disorder.

## 3.2. First Adoption Versus Repeated Amendment

The dataset distinguishes a jurisdiction's initial adoption of a binding legal instrument from later amendments to the same or related legislation. This distinction captures two forms of regulatory development: the spread of regulation to new jurisdictions and the continued revision of existing frameworks.

Of the 151 jurisdictions with at least one binding legal instrument, approximately 85% have adopted a single foundational law that has remained largely unchanged or received only limited revision. A smaller group—including Russia, France, Germany, Singapore, Australia, and the United Kingdom—accounts for a disproportionate share of amendment activity. These jurisdictions have revised their frameworks three or more times, often in response to developments such as the COVID-19 pandemic and the rise of generative AI.

The growth reported in Section 3.1 therefore reflects two parallel processes: initial adoption by previously unregulated jurisdictions and repeated amendment by earlier adopters. Separating these processes provides a clearer measure of regulatory diffusion and avoids overstating the geographic expansion of regulation.

### 3.3. Regional Spread

Regulatory activity has expanded across all UN regional groups, although the timing and pace of adoption have varied. Africa and the Asia-Pacific were the earliest-adopting regions, with the African Group recording the highest proportion of states with at least one coded instrument by 2016. By contrast, the Western European and Other States Group (WEOG) showed relatively limited early activity but expanded rapidly after 2016. This acceleration coincided with major political and technological developments, including the Brexit referendum, the 2016 United States presidential election, the COVID-19 pandemic, and, more recently, the European Union's Digital Services Act and related national measures.

By the end of 2025, Africa and Eastern Europe continued to record the highest proportions of states with regulatory instruments. WEOG, however, had substantially narrowed the earlier gap, indicating greater regional convergence. Responses to information disorder have thus become increasingly global, although important differences remain in when and how quickly jurisdictions adopt regulatory measures.

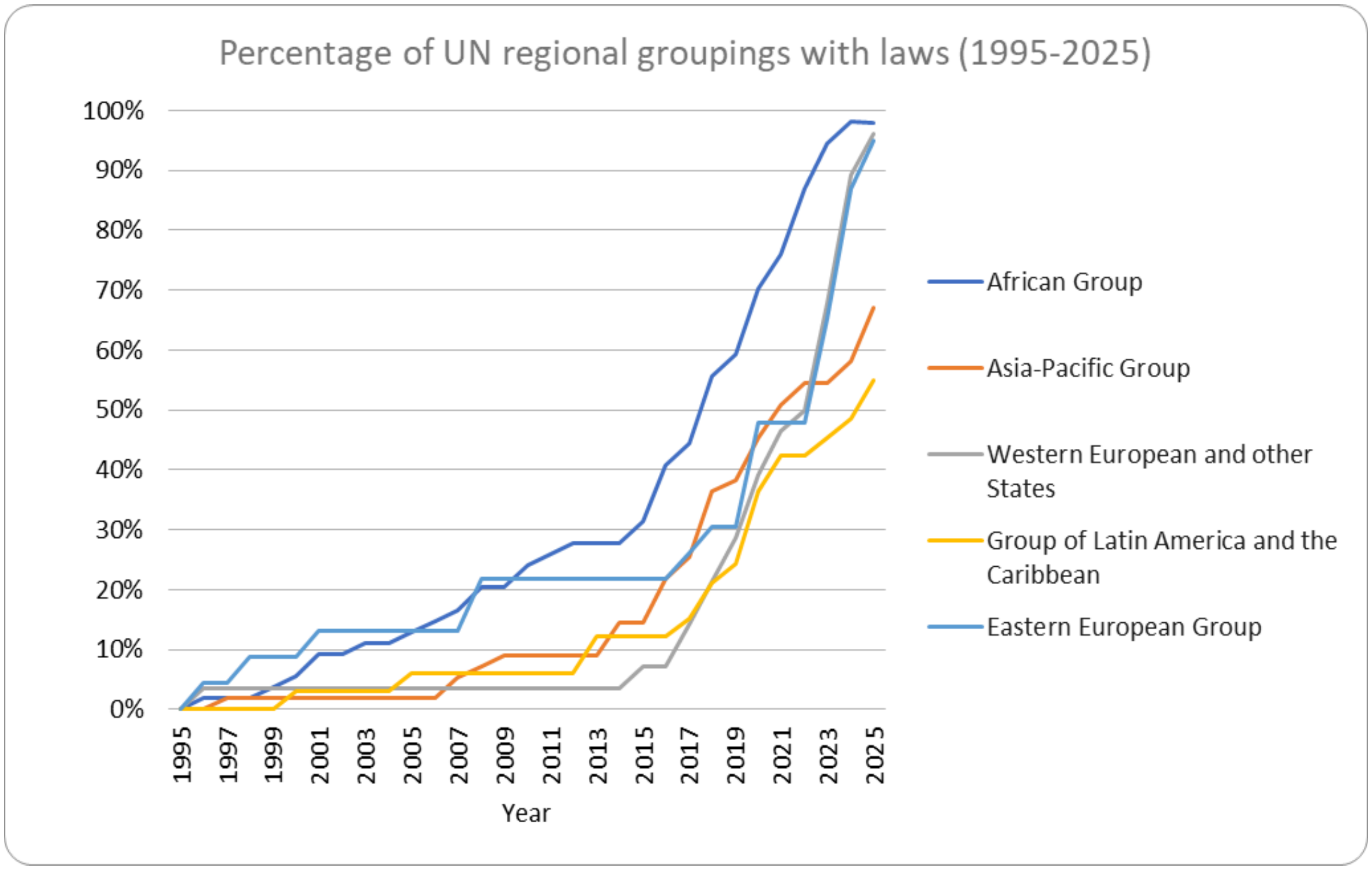

**Figure 3:** Share of states within each UN regional grouping with at least one coded instrument, 1995–2025.

### 3.4. The 2016, COVID-19, and Generative-AI Periods

Three broad phases can be identified in the expansion of regulatory responses to information disorder. The first began around 2016, when the Brexit referendum and the U.S. presidential election intensified public and policy concern about online misinformation and disinformation[6]. The Cambridge Analytica controversy, which became public in 2018, further heightened these concerns.

[6] UNESCO. (2020). Social Media and Elections. Available at https://unesdoc.unesco.org/ark:/48223/pf0000370634.

The second phase emerged during the COVID-19 pandemic in 2020, when regulatory activity increased sharply and the number of measures addressing public-health information roughly tripled.

A third phase began in 2023, alongside the rapid adoption of generative AI. Unlike the earlier phases, which focused largely on false information in general, this period has increasingly addressed specific harms, technologies, and regulated actors.

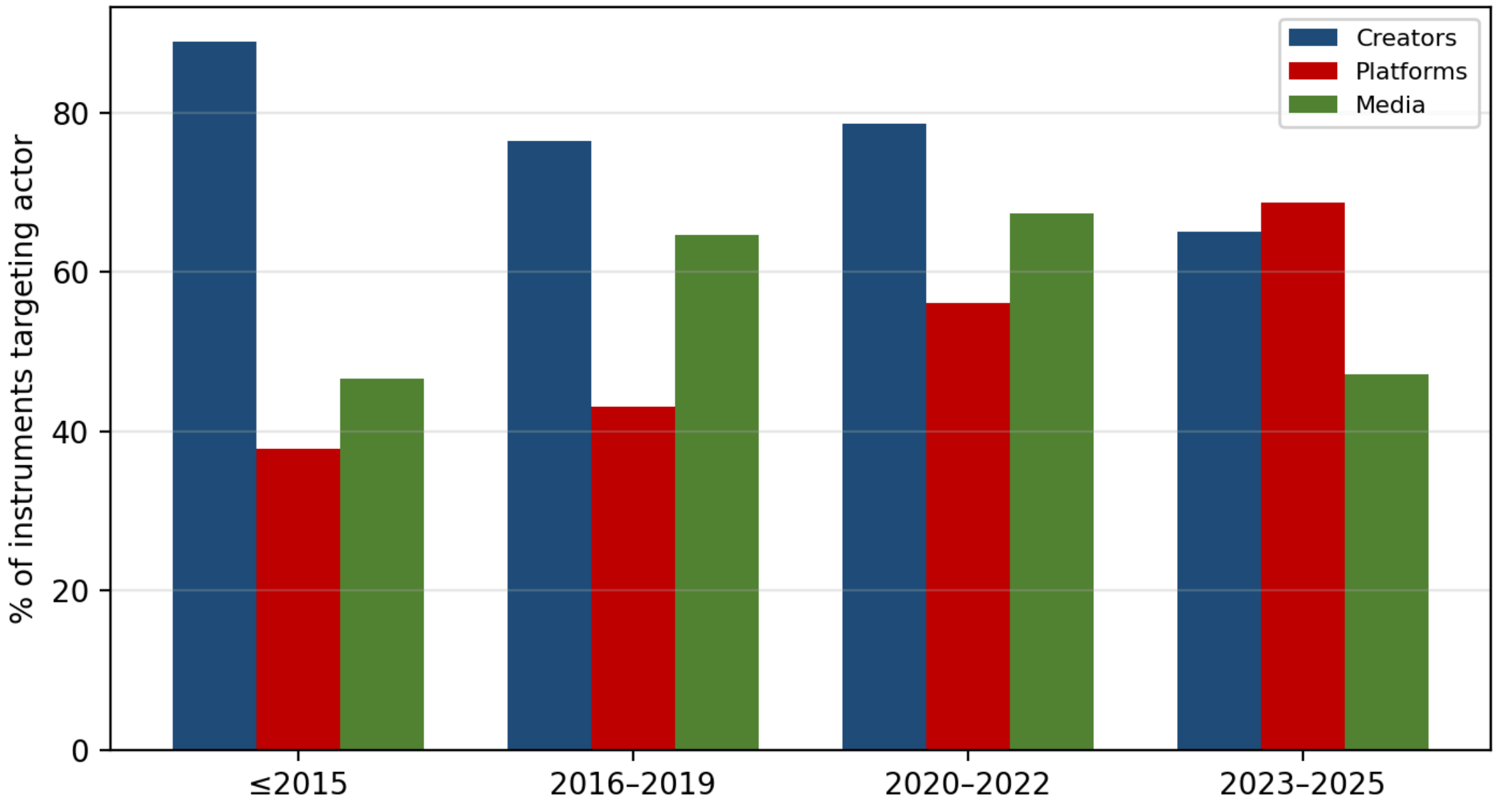

**Figure 4:** Regulatory targets by period (share of instruments naming creators, platforms and media as a regulated actor).

The coded data show this shift clearly. References to deepfakes and AI-generated or synthetic content were almost absent before 2020 but appeared in 17 instruments adopted between 2023 and 2025, representing 9% of instruments from that period. Over the same interval, the share containing election-related provisions increased from 9% in 2020–2022 to 31%, while the share addressing child safety rose from 14% to 50%. Measures concerning impersonation, misuse of digital identity, and non-consensual intimate imagery also emerged primarily after 2023.

Together, these developments indicate a shift from broad prohibitions on false information towards measures targeting specific forms of online harm. South Korea's 2024 amendments[7] addressing deepfake sexual offences illustrate this approach: narrowly defined provisions accompanied by some of the most severe penalties recorded in the dataset.

## 3.5. Regulation in Force and Newly Adopted Measures

Distinguishing instruments currently in force from newly adopted measures clarifies the pattern of recent regulatory growth. Of the 409 instruments in the dataset, 282 remain operative, 85 are no longer in use, and the status of 42 could not be verified. The 85 instruments fall into three categories: 34 binding and soft-law measures that were repealed, withdrawn, superseded, or invalidated; 34 proposals and draft bills that lapsed or were never enacted; and 17 commentary and policy-report entries that were never operative instruments. The latter two categories account for 51 entries that did not confer operative regulatory powers. Only the first category represents measures that ceased to apply, while the remaining categories reflect the dataset's broader coverage of regulatory and policy developments. Of the 34 measures in the first category, 16 addressed public health and were temporary responses to COVID-19 that ended with the emergency. Their expiry therefore does not indicate a broader retreat from regulation.

[7] Act on Special Cases Concerning the Punishment of Sexual Crimes, No. [20459] (2024) (South Korea).

The increase in activity since 2023 is not simply the result of governments replacing expired pandemic measures. Instead, governments have introduced targeted measures addressing specific online harms while retaining earlier frameworks concerned with false information and national security. New measures have therefore largely supplemented, rather than replaced, existing powers, expanding the overall scope of regulation.

### 3.6. Enacted Laws and Proposed Measures

The 45 proposed or draft instruments in the dataset indicate the likely direction of future regulation. Of these, 34 remain under consideration, seven have been enacted, and four have been withdrawn or invalidated. The latter include Australia's Communications Legislation Amendment (Combatting Misinformation and Disinformation) Bill 2024[8], which the government withdrew in November 2024, and Nepal's Social Media Bill[9], introduced in February 2025 and withdrawn in February 2026.

Recent proposals also point to changing regulatory priorities. They increasingly emphasize platform accountability, verification of content origins, and transparency and labelling requirements for AI-generated content, rather than creating new criminal offences for false information. This also signals a shift in regulatory approach, from assessing the truth or falsity of content to targeting specific informational harms, which we discuss in the next section.

This pattern reflects a broader move toward regulating the organizations and technical systems that shape online information. Future measures may therefore focus increasingly on platforms, recommender systems, and AI developers rather than on individual users. Monitoring proposed legislation alongside enacted measures offers an early indication of emerging priorities and should remain part of future annual updates.

## 4. From Falsehoods to Informational Harms

This report outlines various protected interests impacted by MDM, which are now addressed in legislation across jurisdictions. These interests include national security, protection against foreign interference, electoral integrity, public health, and the protection of children and other vulnerable persons. Some associated harms arise from conventional offences but can be amplified by the spread of harmful information, including AI-generated content.

8 Communications Legislation Amendment (Combatting Misinformation and Disinformation) Bill 2024 (Cth).
9 Social Network Bill (2025) (Nepal).

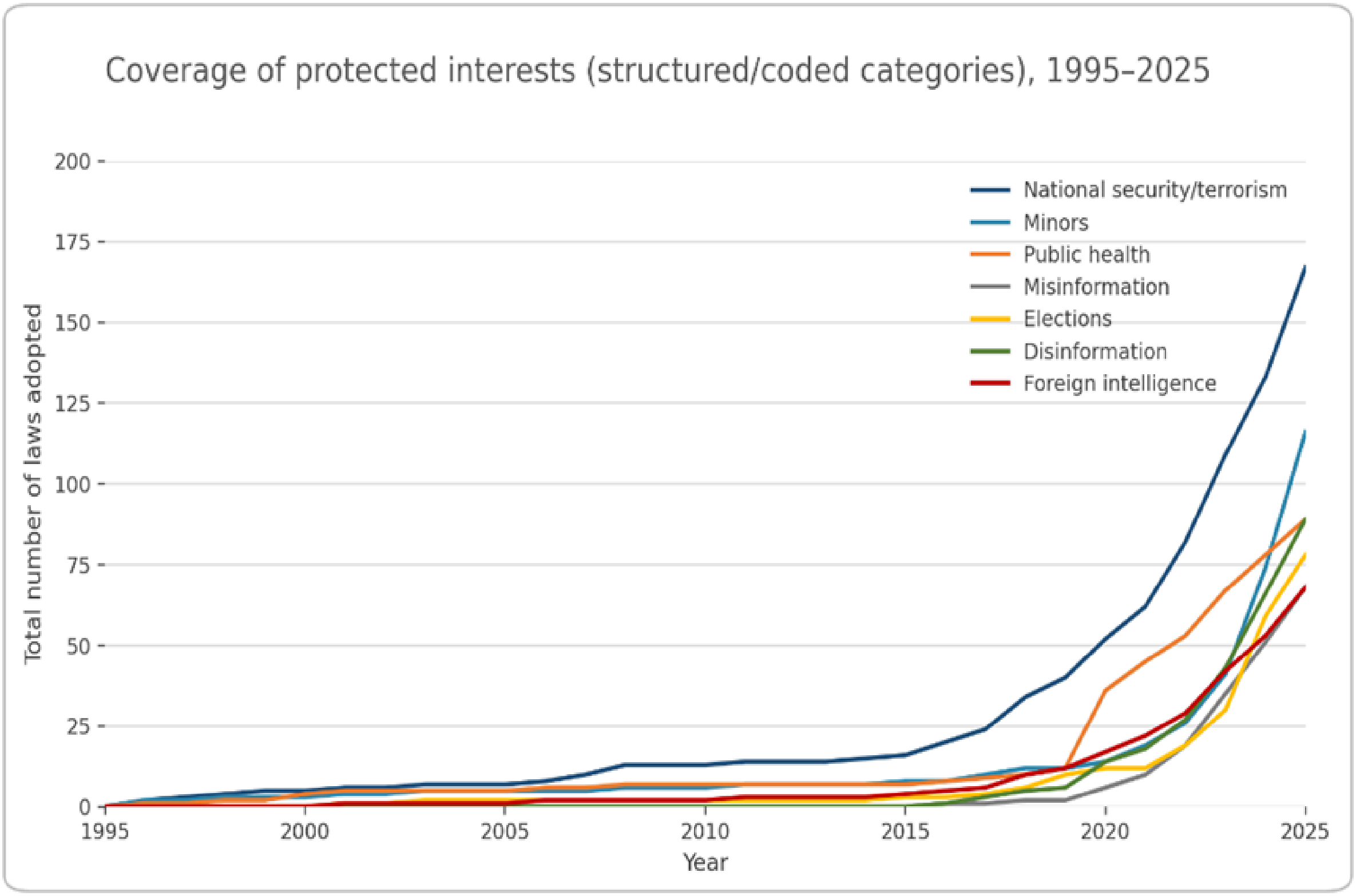

**Figure 5:** Cumulative coverage of protected interests, 1995–2025.

The series are based on the dataset's structured Y/N variables covering misinformation, disinformation, hostile information campaigns/foreign intelligence, national security/terrorism, minors, elections, and public health. At this stage, fraud, privacy, and public order are excluded because they are recorded only in free-text fields and are not independently coded.

## 4.1. National Security and Foreign Interference

Although closely related, national security and foreign interference are divided into two different protected interests in this database. National security includes terrorist activity or activity that threatens public order or sovereignty; foreign interference focuses on deceptive action by external actors seeking to influence social and political conditions in a country.

A longstanding difficulty in defining harm to national security is that it is usually tied to how the government runs the country. This trend of protecting state interests as tools against harmful information that harms national security is climbing steadily. Until 2025, 167 different laws addressed the interests of "national security" or "terrorism" in their legislative provisions. It was constantly in 13 laws passed in 1995, one additional law in 2011, 2014, and 2015. However, there was accelerating momentum: the number of laws grew more than eightfold in nine years, from 20 in 2016 to 167 in 2025.

Foreign interference may be recognized in different terms under legislation, such as "Hostile Information Campaign", "Espionage", and "Counter-Intelligence", which are usually the terms that are commonly used in defining foreign interference under the state legal systems. Other than that, there are also a few other countries that use distinct terms related to national security as the protected interest. For example, Cuba's Decree Law in 2021 on telecommunication and cybersecurity incidents uses "security and internal order of the country".[10]

This limitation of foreign interference is usually related to elections and how MDM affects the voting behaviour during the political contestation. 89 different laws were passed until 2025 that aim to protect against foreign interference. Although the first law that was passed to combat foreign interference was found in 2001, not until 2018 did the battle against foreign entities become popular, which was marked with 10 laws that year. By 2021 and 2023, the number increased to 22 and 42, respectively.

## 4.2. Elections

Although election-related harms overlap with those of foreign interference, countries typically regulate the two separately. Elections as a protected interest tend to be regulated within the scope of how digital propaganda may affect voting behaviours and/or potential voters, and polarize the public due to the differences in political stance and the candidate they support, which eventually may create public unrest during the election period.

The cumulative number of instruments addressing elections rose from 30 in 2023 to 62 in 2024 and 78 in 2025. This increase coincided with widespread distribution of MDM; in addition, 2024 was a "super election" year,[11] as 72 countries held local and national political polls.

Most of these enacted laws aim to protect the integrity of elections, sometimes linked to the right to free and fair elections under Article 25 of the International Covenant on Civil and Political Rights (ICCPR), which protects citizens' right and opportunity to vote and be elected at genuine periodic elections by universal and equal suffrage and secret ballot, guaranteeing the free expression of the will of electors. Some recent laws do not focus only on individual offenders that disseminate false information regarding elections, but also on actions that affect the political environment more generally.[12]

## 4.3. Public Health

The COVID-19 pandemic substantially changed how people around the world interacted from 2020 onwards. Online communication platforms became an increasingly important means of social interaction during periods of physical restriction. Various issues related to the pandemic, ranging from conspiracies about the virus itself to the safety of vaccines and need for mitigating measures, adversely affected public health, also leading some governments to adopt legislation in response.

The cumulative number of instruments addressing public health rose from 12 in 2019 to 36 in 2020 and to 89 by 2025. Some of these laws are still valid, although not actively enforced in practice.

A highlighted example is the Philippines, which passed the Bayanihan to Heal as One Act in March 2020[13]. The Act authorised the President, where the public interest required, to direct the operation of specified privately owned health facilities and public transportation; takeover was permitted only if an enterprise unjustifiably refused or was no longer capable of operating. The law expired in June 2020, and several bills to extend it did not proceed. Five years later, another bill, named the Anti-False Content and Fake News Act,[14] was introduced. This bill proposed to regulate false information in general but included public health as one of the protected interests, under the scope of public interest.

---

[10] Decree-Law No. 35 of 2021 (Cuba), art. 15.

[11] UNDP (n.d.). *A 'Super Year' for Elections*. Available at https://www.undp.org/super-year-elections.

[12] OSCE Office for Democratic Institutions and Human Rights. (2024). *Republic of San Marino, Early Parliamentary Elections: ODIHR needs Assessment Mission Report, 13–15 May 2024*. Available at https://cdn.osce.org/sites/default/files/f/documents/b/1/570300_0.pdf.

[13] Republic Act No. 11469 of 2020 (Philippines).

## 4.4. Children and Vulnerable Persons

Protection of vulnerable groups is also a concern due to the massive spread of MDM. This is related to deepfakes involving vulnerable groups such as minors, women, and persons with disabilities.

MDM has a significant impact on the distribution of intimate image abuse to children. In 2017, 10 laws specifically sought to protect minors from information disorder. The number began increasing to 26 laws by 2022 and 116 such laws in 2025.

Some of these laws that protect minors are related to existing offences that harm children, such as cyberbullying and intimate image abuse, but the recognition of MDM regulation is emerging in various jurisdictions. While some countries adopt the regulation of children's rights in the digital age through digital-resilience policies, such as those adopted by Austria and Belgium (see Section 6.7), other countries are pursuing formal regulation.

Belarus, for instance, enacted the Law on Rights of Children,[15] which regulates children rights in general ranging from child protection, restriction to become worker to family support. This includes protection of children in the country from harmful content on the internet and the imposition of relevant criminal statutes for the offences.

## 4.5. Other Protected Interests

In addition to the harm to national security, elections, public health, and minors, this database also identifies protected interests within the legislation against MDM in the financial sector, public order, and privacy.

As mentioned earlier, while these protected interests are excluded from the analysis in this report, they are identified in the database as free-text fields rather than being independently coded. Nonetheless, a significant number of MDM laws regulating these interests have been recorded, and a more thorough analysis of these interests is planned for future annual updates of the report.

In the financial sector, scams and consumer fraud have been persistent problems since the internet's early days, driving legal systems to adapt. While fraud and forgery were once treated as separate offences, they are now increasingly folded into the broader legal framework for combating MDM. This shift is evident in the dataset, which connects "fraud" and "forgery" to the economic and financial harm caused by MDM, underscoring the financial stakes of the digital information ecosystem. As a result, as of 2025, the database identifies 28 laws that specifically target fraudulent activity as part of efforts to address MDM's impact on the financial sector.

MDM can also directly threaten reputation and privacy. Defamation, for instance, involves false statements that damage a person's standing, and this threat grows as MDM becomes easier to create and spread. In the past, false narratives were typically built by manipulating or recombining existing online information, but modern AI can now fabricate entirely new content from scratch, including sophisticated deepfakes and digital impersonations. These AI-generated deepfakes have become hyper-realistic and hard to detect, spanning techniques such as face and body swapping, facial reconstruction, AI-generated faces, and speech synthesis. Reflecting growing legal awareness of this issue, 38 laws were recorded as of 2025 that explicitly address related concepts including “reputation”, “privacy”, “impersonation”, and “deepfakes”.

[14] Anti-False Content and Fake News Act, S.B. No. 191, 20th Cong. (2025) (Phil.).
[15] Law of the Republic of Belarus No. 86-Z of July 12, 2025, On Amending the Laws on Ensuring Children's Rights (Belarus).

Finally, we turn to the limitations of public order. Within the international human rights framework, state-imposed limits on rights, especially those justified by “national security” or “public order”, are frequently debated, with national security typically concerning threats to state sovereignty and public order concerning large-scale disturbances or damage to public property. The concept of public order has become increasingly contentious amid the global rise of autocracy, as governments use it to justify regulations that claim to curb MDM but ultimately suppress human rights. This broad interpretation has demonstrably harmed free expression and weakened public trust in institutions. Reflecting its legal weight, terms like “public order”, “public nuisance”, and “disturbance” appeared in 91 laws as of 2025.

Among all these limitations on harmful expressions that are regulated within the legislation, targeted actors are another key issue assessed in the database, as different actors play distinct roles in the stages of creating, distributing, and consuming MDM. This is the focus of the next section, which discusses the responsibility of MDM actors.

## 5. Who Bears Responsibility?

The impact of MDM can be tracked through the three phases: creation, dissemination, and consumption. This section focuses on these three phases and identifies various parties who are responsible for the spread of MDM.

A key question is who should be held to account. Traditional internet-related offences are mostly committed by non-state actors such as individuals or organizations. Conduct by state actors may also engage domestic law and, in some circumstances, international legal obligations. However, with the internet and AI, MDM transcends borders, impacting diverse sectors across multiple societies.

The updated database shows an upstream shift in the focus of adopted legislation, broadening its focus to platforms, messaging services, and online intermediaries alongside continued regulation of individual creators and sharers (see Figure 6). Several other parties are also targeted, such as media and political actors, technological corporations, and government institutions.

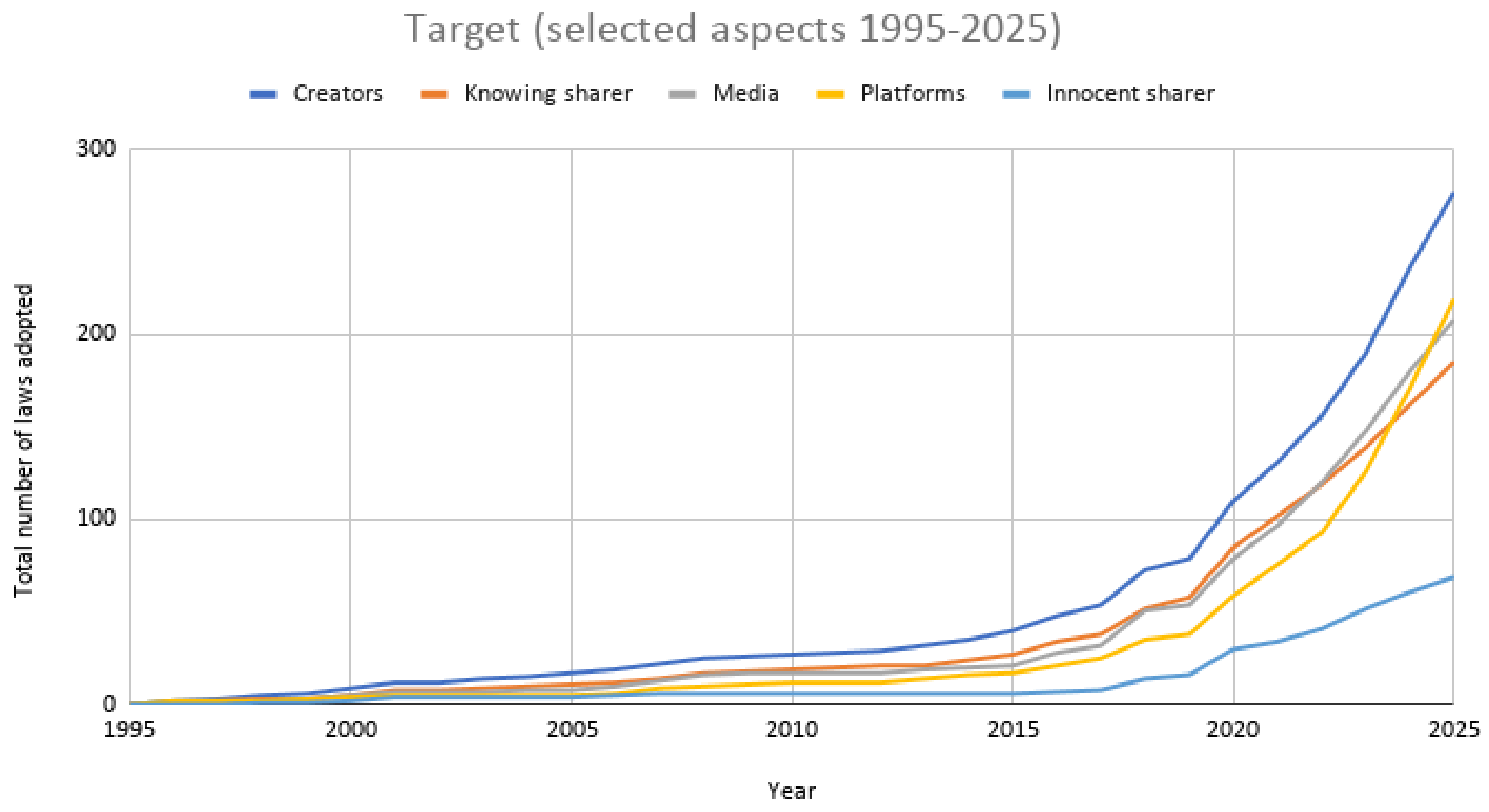

**Figure 6:** Targets of Legislation from 1995-2025.

## 5.1. Creators and Sharers

While creators are defined as individuals who create false information, the dataset distinguishes knowing sharers from innocent sharers. This distinction reflects the earlier definition of MDM illustrated in Table 1, emphasizing that the primary question for legislation should not be whether information is true or false, but whether the information was created or shared with malicious intent. Knowledge that information is false does not necessarily establish an intention to deceive or cause harm; the dataset therefore distinguishes knowledge of falsity from broader questions of motive.

Creators remain the primary target, as indicated in Figure 6. The cumulative number of instruments targeting creators rose from 54 in 2017 to 73 in 2018, 110 in 2020, and 277 by 2025.

The cumulative number of instruments targeting knowing sharers rose from 38 in 2017 to 52 in 2018, 85 in 2020, and 185 by 2025. The COVID-19 pandemic also had a significant impact on the number of laws enacted. In 2020, there was a substantial increase, which climbed from 58 laws to 85 laws. This number remained high until it reached 185 laws in 2025.

The cumulative number of instruments applying to unintentional or careless sharers also increased. Although this type of law is the lowest among the data on the targets of laws, laws targeting innocent sharers have been rapidly growing since 2018 (14 laws), after decades when the laws recorded were under 10 different laws around the world. In 2025, the laws are now 69 laws, but it appears to indicate that various laws passed may rather focus on the large amount of their distribution.

## 5.2. Media and Political Actors

In defining this category, this report focuses on media organizations that share news to the public. This includes traditional media such as television and newspapers that expanded their distribution through online platforms, notably social media, and digital-native media organizations that primarily disseminate news through online platforms.

As seen in the graph, the number of laws adopted is similar throughout the years, and even overlapped in 2022 with the figure of laws targeting knowing sharers, with 120 laws. By 2025, the cumulative number of instruments targeting media organisations had reached 208, making media the third-most frequently targeted actor in the dataset. These types of laws are concerning due to the harm to freedom of the press and freedom to access information. This was indicated, among others, in Kyrgyzstan's 2025 Code of the Kyrgyz Republic on Offences, which penalizes unreliable (false) information with use of mass media or the Internet[16].

In a notable 2025 ruling, the European Court of Human Rights (ECHR) adjudicated the case of *Bradshaw and Others v. the United Kingdom*[17]. The plaintiffs, a group of former UK Members of Parliament, claimed the government had failed to adequately investigate and counter Russian disinformation campaigns that allegedly targeted the 2014 Scottish Independence referendum, the 2016 Brexit referendum, and the 2019 general election. The ECHR found that the United Kingdom had not violated Article 3 of Protocol No. 1 to the European Convention on Human Rights, which protects the right to free elections. While the Court acknowledged the reality of foreign influence in elections, it concluded that the UK's response had been sufficient. The ruling cited several countermeasures, including parliamentary inquiries, the creation of specialized government units, and the passage of significant new legislation like the Elections Act 2022, the National Security Act 2023, and the Online Safety Act 2023. The case is significant because the Court considered the state's positive obligations to protect electoral processes from sufficiently serious foreign interference, while finding no violation of the facts. Crucially, the ruling makes it imperative to observe how national legislatures will incorporate its principles. Future legislation will need to address key areas, particularly the protection of electoral integrity against interference and the delineation of state obligations regarding digital information.

## 5.3. Platforms and Messaging Services

The number of laws targeting platforms follows a trend similar to that of laws targeting creators and sharers. In recent years, platforms have faced substantial scrutiny over how responsible and liable they should be for managing the algorithms that drive the massive spread of MDM to internet users. Notable examples include Meta's role in amplifying disinformation against the Rohingya in Myanmar and the tendency of generative-AI systems to produce false or misleading outputs, often described as "hallucinations".

The number of laws adopted that regulate and sanction platforms is increasing, although it remains lower compared to other actors in this dataset. After a significant increase in 2020 to 59, from 38 laws in 2019, this number climbed steadily and reached 219 in 2025, exceeding the 185 recorded for knowing sharers.

Prior to 2025, several cases concerning online platform liability were tried in the United States. While not focused specifically on misinformation and disinformation, these cases signaled a similar pattern of a potential shift in legal discourse on online platform responsibility.

[16] Code of the Kyrgyz Republic on Offences, art. 109-3 (amended 2025) (Kyrgyzstan).
[17] Bradshaw and Others v. The United Kingdom, App. No. 15653/22 (2025).

An example is *NetChoice, LLC v. Paxton*[18], which involved a challenge to Texas House Bill 20. This law sought to prohibit large social media platforms from engaging in viewpoint-based content moderation (except for illegal content) and required them to disclose their moderation practices. The plaintiffs, NetChoice and the Computer & Communications Industry Association, argued the bill was unconstitutional under the First Amendment. The Supreme Court ultimately nullified the lower court's ruling and remanded the case back to the lower courts for further consideration of the First Amendment challenge.

This procedural outcome mirrors the decision in *Gonzalez v. Google LLC*[19]. In that case, a family contended that Google, through YouTube's recommendation algorithms, was liable for promoting terrorist content that led to their child's death. The Supreme Court also vacated the lower court's decision and remanded the case for reconsideration.

Although these early cases did not rule directly on the issue of false information, they marked a move toward scrutinizing the legal responsibilities of online platforms for harmful content, broadening the focus beyond individual liability. Recent decisions in 2026 have tested platform liability for harms to young users (Section 6.2) and for AI-generated false statements (Section 6.4).

Together, all the regulated interests intended to prevent harm caused by MDM, along with the actors who should be held responsible, point to a significant question: how should laws and policies intervene on the issue of MDM? The next section discusses the findings of this study, which categorizes the various approaches that legislation has taken thus far.

# 6. The Regulatory Toolbox

Given the stages of creation, distribution, and consumption in the digital information ecosystem, adequate legislative responses become crucial to intervening in information disorder at each of these phases. This section discusses the trend of legislative approaches taken by countries around the world. As identified in the database, there are different types of instruments, including hard and soft law. In addition, as discussed in the previous section, these interventions range from the creation of false information, dissemination through platforms, punitive sanctions, as well as literacy policy and content labelling. Instead of targeting the feared harm caused by MDM, legislation often targets content related to the adjacent harm, which justifies the discussion of such legal interventions in the pipeline of production, distribution, and consumption of digital information.

## 6.1. Production and Synthetic-Content Controls

Regulations towards private activities that may cause harm to other parties remain part of legal intervention. Australia has proposed stricter controls. The Criminal Code Amendment (Using Technology to Generate Child Abuse Material) Bill 2025 would create new offences for generating child abuse material using a carriage service.[20] Canada's Protecting Victims Act, assented to in 2026, amended the existing mandatory-reporting regime for Internet child sexual abuse and exploitation material, including by broadening the definition of Internet services and revising notification and related duties.[21]

[18] *Moody v. NetChoice, LLC* and *NetChoice, LLC v. Paxton*, 603 U.S. 707 (2024).
[19] *Gonzalez v. Google LLC*, 598 U.S. 617 (2023).
[20] Criminal Code Amendment (Using Technology to Generate Child Abuse Material) Bill 2025 (Cth).
[21] *Protecting Victims Act*, S.C. 2026, c. 19 (Can.).

Belgium also implemented interventions to protect children, taking a layered approach to creating a safer digital space for children. This includes a law on intimate image abuse against children in 2020[22] and a new criminal law for sexual acts.[23] In 2025, Belgium also planned to introduce digital wallet identification for children,[24] after the national ID was implemented a year before. This will include an age-verification step, which has been under scrutiny regarding the issue of privacy[25]. These plans follow similar steps taken by many countries on age-verification and limiting social media usage for children, such as in Australia,[26] Brazil,[27] Indonesia,[28] and the United Kingdom.[29]

Other than intimate image abuse on children, a greater focus on individuals' privacy has also become a serious consideration by several countries. Denmark is an example of this, with proposals to amend copyright law to protect individuals against unauthorized sharing of digital imitations of their appearance and voice.[30] A similar approach is also taken by Peru, where synthetic pornography that is considered a deepfake and defamation is regulated under Law No. 32314, specifically in the use of AI as a tool to aggravate criminal offences.[31]

## 6.2. Publication and Speaker Liability

Since AI chatbots were launched and became popular among internet users, the creation of false information is massively increasing. This is also exacerbated by AI-embedded features and tools in various online platforms.

The significant increase in the ability to create and share synthetic content raises the long-term question of internet neutrality and how to regulate online platforms. Determining whether online intermediaries should be treated as distributors of the harmful content, or publishers of such online information, becomes crucial, as each designation carries different legal obligations.

In the United States, Section 230 of the Communications Decency Act of 1996 generally protects online services from being treated as the publisher or speaker of content supplied by another information provider. In early 2026, several landmark rulings were issued concerning how Meta, as a social media company, had designed an addictive platform that harms mental health, especially young children. These landmark decisions contributed to a deeper discussion regarding the implementation of Section 230.

---

[22] *Wet ter bestrijding van de niet-consensuele verspreiding van seksueel getinte beelden en opnames* [Law of 4 May 2020 aimed at combating the non-consensual dissemination of images and recordings of a sexual nature], Belgisch Staatsblad [B.S.], June 18, 2020 (Belg.).

[23] *Wet van 21 maart 2022 houdende wijzigingen aan het Strafwetboek met betrekking tot het seksueel strafrecht* [Law of 21 March 2022 modifying the Criminal Code with regard to sexual penal law], Belgisch Staatsblad [B.S.], April 30, 2022 (Belg.).

[24] Borak, M. (2024, May 15). *Belgium launches national digital identity wallet*. Biometric Update. https://www.biometricupdate.com/202405/belgium-launches-national-digital-identity-wallet.

[25] UNICEF. (2025). *Drawing a line in digital spaces: Age-based restriction of social media* [Policy Note]. 2.

[26] Online Safety Amendment (Social Media Minimum Age) Act 2024 (Cth) (Austl.).

[27] *Lei No. 15.211 de 17 de setembro de 2025, Estatuto Digital da Criança e do Adolescente* [Law No. 15,211 of September 17, 2025, Digital Statute for Children and Adolescents], Diário Oficial da União [D.O.U.], September 18, 2025 (Braz.).

[28] Peraturan Pemerintah No. 17 Tahun 2025 tentang Tata Kelola Penyelenggaraan Sistem Elektronik dalam Perlindungan Anak [Government Regulation No. 17 of 2025 on the Governance of Electronic System Operation in Child Protection], Lembaran Negara Republik Indonesia [L.N.R.I.] No. 36, Tahun 2025 (Indon.).

[29] Online Safety Act 2023, c. 50 (UK).

[30] Forslag til Lov om ændring af lov om ophavsret (Indførelse af en præstationbeskyttelse og beskyttelse mod digitalt genererede efterligninger mv.) [Proposal for an Act to Amend the Copyright Act (Introduction of performance protection and protection against digitally generated imitations, etc.)] KUU Alm.del - Bilag 232, Folketinget 2024-2025 (Den.).

[31] Ley que modifica el Código Penal, Decreto Legislativo 635, y la Ley 30096, Ley de Delitos Informáticos, para incluir el uso de la inteligencia artificial en la comisión de delitos [Law that amends the Criminal Code, Legislative Decree 635, and Law 30096, the Computer Crimes Law, to include the use of artificial intelligence in the commission of crimes]. (2025). Congreso de la República del Perú [Congress of the Republic of Peru].

As stated in the methodology section, this report's primary focus is on national regulatory frameworks, as systematic coverage of subnational measures is not feasible. However, while the rulings are subnational, they serve as a crucial illustrative case study of the national-level regulatory tensions discussed throughout this report. The court's direct engagement with the limitations of Section 230 provides a concrete example of the very challenges facing federal policymakers. By examining the court's reasoning, we can better understand the practical hurdles and interpretive battles that define the broader U.S. approach to regulating information disorder.

In several jurisdictions, criminal offences have been formulated against individuals, even without proof of malicious intent, especially during the COVID-19 pandemic. This is evidenced by the amendment to Uzbekistan's Criminal Code in 2020, which introduced criminal punishment for the dissemination of false information. Similarly, the 2022 Ugandan Computer Misuse (Amendment) Act[32] criminalizes the distribution of false information, with a specific provision if the offence is committed by the alleged perpetrator under disguise or false identities on social media. (In March 2026, Uganda's Constitutional Court declared the 2022 Amendment Act null and void because Parliament had failed to ascertain the required quorum; it separately held criminal-libel provisions unconstitutional.[33])

Despite the emerging laws on publisher and speaker's liability, other states also implement limitations of these punitive approaches. The United States passed the Take It Down Act in 2025[34], which includes good-faith liability protections and a severability clause. This also occurred in Kenya, in the form of a policy issued by the Media Council[35] that has a comprehensive ethical code integrating AI governance, platform liability, and content moderation duties.

## 6.3. Distribution, Amplification, and Virality

The democratization of information production and consumption has empowered the public to create various forms of online information. This has caused a massive flow of digital information because anyone can create and publish content, which in turn makes distinguishing accurate and credible information increasingly challenging. Moreover, this decentralized model of information creation can bypass traditional editorial gatekeeping, particularly amid the competition to produce sensational, attention-grabbing material that drives up readership and advertising income. As a result, MDM is amplified by this model of online information creation.

MDM can spread rapidly in the digital environment, shaped not only by the nature of harmful content but also by how platforms amplify its reach. However, considering the protected interests discussed in Section 4, mere gossip, rumour, or viral false information on the internet should not be treated as unlawful expression unless such information impacts those interests.

---

[32] The Computer Misuse (Amendment) Act, 2022 (Uganda).
[33] Alternative Digitalk Limited and Others v The Attorney General (Consolidated Constitutional Petitions No. 34, 37 & 42 of 2022) [2026] UGCC 3 (17 March 2026).
[34] TAKE IT DOWN Act, Pub. L. No. 119-12, 139 Stat. 55 (2025).
[35] The Media Council (Code of Media Practice), 2025 (Kenya).

Countries have adopted various approaches to address this problem. Nepal enacted its first MDM-related provisions in the 2017 National Penal Code.[36] Although the government subsequently proposed a new social media bill, it was withdrawn after widespread protests against corruption and proposed social media restrictions, amid shifting political dynamics.

In contrast, Kyrgyzstan has pursued incremental legal reform. The country amended its Penal Code in 2021 to criminalize the dissemination of false or unreliable information via mass media and the internet[37].

## 6.4. Platform Governance

The US cases discussed earlier concerned platform liability for harmful content rather than MDM specifically.

Interestingly, in a recent case regarding false and defamatory statements, the Munich Regional Court I, in preliminary-injunction proceedings, held that Google could be directly liable for false statements generated in its AI Overview feature.[38] The decision by the Munich Regional Court (a preliminary injunction, which is now being appealed by Google) referenced the EU's Digital Services Act. The court treated the AI-generated summary as content attributable to Google rather than merely as a reproduction of third-party search results.[39]

Similarly, in 2024, Ireland passed the Online Safety Code.[40] This regulation mandates platform safety obligations for harmful online video content. Singapore's Online Safety (Miscellaneous Amendments) Act 2022, which took effect in February 2023, strengthened platform accountability, redress for victims, and enforcement powers.

These examples indicate that mere sanction is not the only approach or tool to shut down harmful content, especially given that individuals are no longer the only party that can be liable. Platforms amplify and, increasingly, are being recognized themselves as creators of false and/or harmful information.

## 6.5. Labelling, Provenance, and Authentication

A different approach in implementing legal and policy interventions is also emerging in determining the origin of content, whether it is created by humans or AI-generated. The label on such content aims to communicate to internet users the authenticity of the content that is distributed. Particularly related to social and political issues, such as public protests or elections, knowing that content is AI-generated may encourage users to assess its context and reliability before sharing it.

Governments in several jurisdictions have adopted such policies. China's 2025 *Measures for the Labeling of AI-Generated Synthetic Content* apply to text, images, audio, video, virtual scenes and other AI-generated or synthesised information. The measures require service providers to use explicit user-visible labels and/or implicit metadata labels; sanctions may arise under other applicable laws.[41]

[36] National Penal Code (Muluki Criminal Code) Act, 2074 (2017) (Nepal).
[37] Кыргыз Республикасынын Кылмыш-жаза кодекси [Criminal Code of the Kyrgyz Republic], Law No. 127 (2021) (Kyrg.).
[38] Landgericht München I [LG München I] [Regional Court of Munich I] May 28, 2026, Case No. 26 O 869/26 (Ger.).
[39] Digitale-Dienste-Gesetz [DDG] [Digital Services Act], Bundesgesetzblatt [BGBl.] I (2024) (Ger.).
[40] Coimisiún na Meán, Online Safety Code (2024) (Ir.).

The labelling approach is also currently planned by France in a bill to identify AI-generated content on the internet.[42] Similarly, Romania is also drafting a law on labelling content as a response to the deepfake phenomenon in 2023.[43]

## 6.6. Correction, Removal, and Redress

Content labelling represents an emerging approach that could help prevent the dissemination of false information. However, this raises the question of what interventions are available for false information that has already been disseminated. Beyond punitive sanctions, legal frameworks also grant authorities the power to remove content.

Within this broader pattern, the specific choice of decision-maker still varies across jurisdictions. In determining the truth of information, some countries empower existing government agencies, others establish new bodies for the purpose, and still others assign the task to courts or judges.

Brazil, for example, through its 2019 Resolution,[44] granted power to its Electoral Court for electoral propaganda and false electoral information. In Latvia, the National Electronic Mass Media Council (NEPLP), an independent media regulator operating under the Electronic Mass Media Law, supervises electronic media and may restrict services or content that threaten national security or public order. Its recent strategy places particular emphasis on protecting Latvia's information space, including in response to Russian propaganda and disinformation.

Legal frameworks in many states emphasise that determining the truth of disputed information is currently handled through judicial processes by courts and other judicial bodies. However, some of these same laws still grant executive government authorities the power to make such determinations.

## 6.7. Media Literacy and Resilience

The interventions discussed in previous subsections address the production and distribution of harmful information. However, addressing consumption is equally important, as it involves empowering citizens to think critically about online content and develop digital literacy and resilience. The scope of media literacy has therefore broadened to include digital literacy and AI literacy. Many countries have implemented educational policies reflecting this expanded approach.

---

[41] 人工智能生成合成内容标识办法 [Measures for the Labeling of AI-Generated Synthetic Content] (2025) (China).
[42] Proposition de loi visant à identifier les images générées par intelligence artificielle publiées sur les réseaux sociaux [Bill to Identify Artificial-Intelligence-Generated Images Published on Social Networks] (2024) (Fr.).
[43] Lege privind utilizarea responsabilă a tehnologiei în contextul fenomenului deepfake [Law on the Responsible Use of Technology in the Context of the Deepfake Phenomenon], PL-x No. 471/2023 (Rom.).
[44] Resolução No. 23.610 de 18 de dezembro de 2019 [Resolution No. 23,610 of December 18, 2019], Diário Oficial da União [D.O.U.], (Braz.).

In Austria, there are two key frameworks about digital resilience, namely the Austrian Youth Strategy[45] and the Digital Austria Act.[46] The first focuses on policy to prepare young people in a digital environment. Meanwhile, the Digital Austria Act addresses digitalization issues including digital skills.

A narrower approach to combating disinformation was adopted by Sweden with the establishment of the Psychological Defence Agency in 2022. Initially under the Ministry of Justice, the Agency moved to the Ministry of Defence in 2023. Its mandate focuses on identifying and analysing malign foreign information influence directed at Sweden or Swedish interests, coordinating responses, and strengthening public resilience through education and support.

Misinformation and disinformation remain among the leading short-term global risks identified in the World Economic Forum's 2026 survey. Punitive sanctions are unlikely to succeed as a standalone intervention. Just as governments rarely rely on a single regulatory approach (discussed in Section 7), an effective response to MDM should combine punitive measures with education-focused policy and cross-sector cooperation, rather than treating enforcement of punishment as a self-sufficient solution.

# 7. Regulatory Models and Regional Patterns

The preceding sections examined the growth, geographical spread, and substantive focus of responses to information disorder. This section examines the approaches taken by governments to regulate information disorder. Although jurisdictions may pursue similar objectives, they often adopt different legal approaches because of differences in constitutional traditions, institutional capacity, policy preferences, and perceptions of online risk. Examining regulatory design is therefore as important as counting the instruments adopted.

Governments rarely rely on a single approach. Many combine criminal punishments, platform obligations, administrative powers, individual remedies, and transparency requirements within the same framework. This section applies a six-model typology to examine how these approaches vary across jurisdictions and regions and how frequently they are combined. Because the models overlap, a single instrument may be assigned to more than one category.

## 7.1. An Empirical Typology

Instrument counts indicate the scale of regulatory activity but do not show how governments intervene. To support comparison across jurisdictions and over time, binding legal instruments were classified according to their substantive features rather than their formal legal designation. The first five models apply to binding legislation, while the sixth captures non-binding policy and soft-law measures. Table 3 presents their definitions, examples, and approximate prevalence.

| Model | Operational definition from coded variables | Illustrative instruments | Approx. share of hard-law instruments |
|---|---|---|---|

[45] Bundeskanzleramt Österreich [Federal Chancellery of Austria]. (n.d.). *Österreichische Jugendstrategie* [Austrian Youth Strategy], Available at https://www.bundeskanzleramt.gv.at/en/agenda/youth/austrian-youth-strategy/.

[46] Bundesministerium für Finanzen [Federal Ministry of Finance]. (n.d.). *Digital Austria Act*. Available at https://www.digitalaustria.gv.at/.

| | | | |
|---|---|---|---|
| Truth-policing | Empowers a designated entity to determine truth/falsity of a claim (Y), typically paired with correction or removal powers | Singapore POFMA (2019); Code of administrative offences of the Russian Federation No. 195; Egypt Penal Code arts. 80C/102bis | ~25% |
| Criminal harm prevention | Imposes fines or imprisonment on those who create or knowingly disseminate prohibited content, without necessarily granting a public authority correction or removal powers. | False-information provisions in penal codes across Africa, the Middle East, and parts of Asia; criminal offences covered by Germany's NetzDG framework | ~45% |
| Platform systems governance | Imposes procedural or organizational duties on platforms, including risk assessment, transparency, reporting, and notice-and-action requirements. | EU Digital Services Act; UK Online Safety Act (2023); Australia's (withdrawn) misinformation bill (2024) | ~35% (rising sharply since 2022) |
| Victim-centred rights and remedies | Addresses harms to identifiable individuals, including intimate-image abuse, impersonation, and harmful deepfakes, and provides criminal or civil remedies. | UK Online Safety Act (2023); South Korea's 2024 deepfake sexual-offence amendments; U.S. state laws on non-consensual intimate imagery | ~6% (concentrated 2023–2025) |
| Provenance regulation | Requires labelling, watermarking, or disclosure of AI-generated or synthetic content, or verification of platform or user identity. | China's Provisions on the Administration of Deep Synthesis 2023; Taiwan advertisement-labelling amendment; EU AI Act transparency obligations | 4%, almost entirely adopted after 2023 |
| Resilience-based approaches | Soft-law and policy measures focused on media literacy, digital citizenship, or institutional coordination rather than binding obligations. | Nordic hybrid-threat and psychological-defence frameworks; national media-literacy action plans; multi-stakeholder codes of practice | n/a – corresponds closely to the 56 soft-law/policy instruments in Table 2 |

**Table 3:** Six-model typology of regulatory approaches in the coded dataset.

**Note:** Percentages are approximate. The categories overlap because many instruments combine several approaches. A single instrument may therefore be counted under more than one model. Soft-law and policy measures are reported separately under the resilience and literacy model.

Two findings emerge from this classification. First, the models represent approaches that are frequently combined rather than distinct alternatives. For example, the EU Digital Services Act and UK Online Safety Act combine platform obligations with transparency requirements, while South Korea's deepfake legislation combines criminal sanctions with remedies directed at affected individuals.

Second, official truth determination and criminal harm prevention remain among the most common approaches used in binding legal instruments. This reflects the perspective that criminal law is commonly employed against a range of offences that cause harm, involving the characterization of such conduct as harmful or as carrying a risk of harm[47]. However, the tension between free expression and the complexity of legal liability, as this study has shown, may encourage further assessment and a rethinking of criminal law's dominant approach.

Since 2023, however, much of the growth has occurred in platform governance, individual rights and remedies, and content-authenticity rules. These newer approaches have generally supplemented rather than replaced earlier forms of regulation, producing broader and more layered frameworks.

## 7.2. The Upstream Shift, Restated by Model

The trends identified in Section 3.4 become clearer when examined through the six-model typology. Platforms were regulated in 38% of instruments adopted before 2016, compared with 69% of those adopted between 2023 and 2025. Over the same period, the proportion targeting individual creators declined from 89% to 65%, while the proportion targeting known disseminators fell from 64% to 35%. These changes suggest the growing importance of platform governance and content-authenticity rules alongside continuing criminal sanctions. Individual creators remain the most frequently regulated actors across the full dataset, appearing in 300 of 409 instruments, but recent measures increasingly impose obligations on platforms as well as, or instead of, individuals.

This pattern indicates a broader shift from regulating individual conduct toward regulating the digital systems through which content is created, distributed, and amplified. The European Union's Digital Services Act and the United Kingdom's Online Safety Act illustrate this approach by imposing extensive duties on large online platforms. Similar features appeared in Australia's withdrawn 2024 misinformation bill and in several measures introduced across Africa and the Asia-Pacific region between 2024 and 2025. Platform governance is therefore becoming an increasingly prominent, although not exclusive, component of regulatory responses to information disorder.

## 7.3. Regional and Regime-type Patterns

The dataset shows substantial regional variation in both the number of instruments adopted and the powers they confer. Figure 7 compares the mean Composite Power Index[48] across UN regional groups.

---

[47] McNamara, L., Quilter, J., Hogg, R., Douglas, H., Loughnan, A., & Brown, D. (2018). Theorising criminalisation: The value of a modalities approach. *International Journal for Crime, Justice and Social Democracy, 7*(3), 91–121. DOI: 10.5204/ijcjsd.v7i3.511.

[48] See below, Section 8.6.

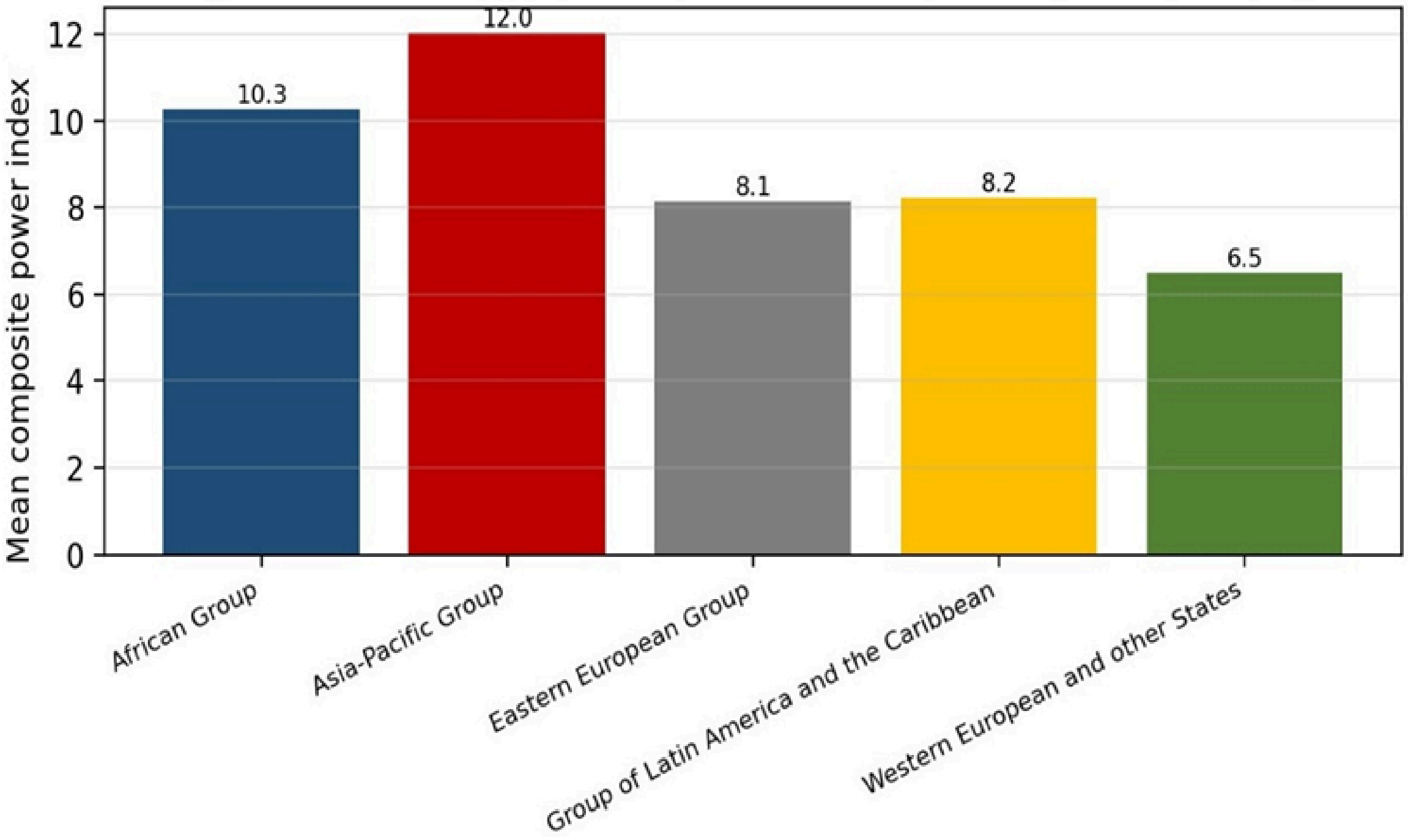

**Figure 7:** Mean Composite Power Index by UN regional grouping across all coded hard-law and soft-law instruments.

The Asia-Pacific Group has the highest mean composite power index score (12.0), followed by the African Group (10.3), while WEOG has the lowest (6.5). These figures represent regional averages rather than maximum scores. The index is the sum of the coverage variables and the weighted powers, giving a range of 0 to 20. Regional differences therefore reflect not only the volume of regulation but also the breadth and severity of the powers granted. Despite considerable variation within the Asia-Pacific region, its coded instruments confer broader regulatory powers, on average, than those in other regions.

The correlation results provide further context but do not imply causation. Government effectiveness has almost no association with the Composite Power Index ($r = -0.03$), while the rule of law has only a weak negative correlation ($r = -0.26$). Regulatory strength therefore cannot be explained by either measure alone. Constitutional arrangements, domestic priorities, and concerns about foreign interference may also influence regulatory design, although these relationships cannot be tested with the present dataset and would require further qualitative assessment of the legislation's efficacy.

Legal traditions may provide an additional explanation, but the patterns are not uniform. Some common-law jurisdictions, particularly Singapore and India, empower administrative authorities to issue correction or removal directions, while the United Kingdom and Australia have placed greater emphasis on platform duties. Civil-law jurisdictions in continental Europe generally define regulatory obligations in detailed legislation and distribute decision-making among courts and independent regulators. Broad false-information offences remain comparatively common in parts of Eastern Europe, particularly in legal systems influenced by Soviet or Russian law, sometimes with limited procedural safeguards. These differences suggest that regional patterns reflect distinct combinations of legal tradition, institutional structure, and political context.

### 7.4. What Withdrawn, Repealed and Struck-down Measures Reveal

The dataset includes 34 instruments that were withdrawn, repealed, or invalidated by a court. Although these instruments constitute a small share of the dataset, they help identify the conditions under which regulatory measures face legal or political resistance. They also provide an indication of the factors that may affect whether particular approaches endure. Temporary measures that expired after the COVID-19 emergency should, however, be distinguished from instruments rejected on constitutional or political grounds.

Several post-2023 cases involving official determinations of truth have faced constitutional review. In 2024, the Bombay High Court invalidated provisions authorizing an official fact-checking unit to identify false or misleading information concerning the Indian government.[49] Mongolia's Constitutional Court similarly invalidated the country's criminal false-information provision in November 2025.[50] Although the number of cases is too small to establish a general pattern, these decisions indicate that measures granting public authorities broad powers to determine the truth of contested claims may face heightened scrutiny, particularly when they lack clear procedural safeguards or disproportionately restrict freedom of expression.

No platform-governance instrument coded in the dataset had been judicially invalidated by the end of the study period. Challenges to this approach have instead been primarily political and practical. Australia's Communications Legislation Amendment (Combatting Misinformation and Disinformation) Bill 2024 was abandoned following sustained political opposition and concerns about freedom of expression. The implementation of the European Union's Digital Services Act, meanwhile, illustrates the institutional and technical demands involved in supervising large platforms and enforcing system-wide obligations. These cases suggest that regulatory approaches encounter different constraints: official truth-determination measures may be especially vulnerable to constitutional challenge, whereas platform governance is more often shaped by political acceptability and enforcement capacity.

## 8. Powers, Safeguards and Institutional Design

The regulatory models examined in Section 7 identify the main approaches governments use to address information disorder. They do not, however, show how authority is allocated or exercised. Instruments within the same model may confer different powers, assign responsibility to different institutions, and provide different safeguards. These features influence both the extent of regulatory intervention and its operation in practice.

This section examines four aspects of regulatory design: enforcement powers, institutional arrangements, procedural safeguards, and accountability mechanisms. It then introduces the Composite Power Index, which summarises the powers associated with each instrument and supports comparison across jurisdictions. Alternative weighting schemes and sensitivity tests are used to assess whether the resulting patterns remain consistent.

---

[49] The Hindu. (2024). *Bombay HC Formally Strikes D"wn Centre's Fact-Check Un"t, Calls Amended IT Rules Unconstitutional.* The Hindu, available at https://www.thehindu.com/news/national/bombay-hc-formally-strikes-down-centres-fact-check-unit-calls-amended-it-rules-unconstitutional/article68684934.ece.

[50] International IDEA. (2025). *"Mongolia — November 2025," The Global State of Democracy Democracy Tracker*, Available at https://www.idea.int/democracytracker/report/mongolia/november-2025.

## 8.1. Severity and Breadth of Powers

The reach of a regulatory instrument depends not only on the regulatory approach it adopts but also on the powers it authorises. These range from mandatory corrections and content removal to financial penalties and imprisonment. Examining these individual powers therefore provides a more direct indication of the breadth and severity of regulatory intervention than regulatory models alone. Table 4 reports the principal powers coded in the dataset.

| Power | n | % of 409 instruments |
|---|---|---|
| Power to fine | 287 | 70% |
| Power to imprison | 242 | 59% |
| Power to require removal | 169 | 41% |
| Power to require correction | 107 | 26% |
| Empowers an entity to determine truth/falsity | 104 | 25% |

**Table 4:** Powers coded across the 409 instruments adopted or proposed between 1995 and 2025.

**Note:** Categories are not mutually exclusive. Percentages are calculated using all 409 instruments as the denominator.

The dataset shows that coercive powers remain the most common feature of regulatory frameworks addressing information disorder. Fines appear in 287 instruments (70%), while 242 instruments (59%) provide for imprisonment. Content-removal powers appear in 169 instruments (41%), mandatory corrections in 107 (26%), and official truth-determination powers in 104 (25%).

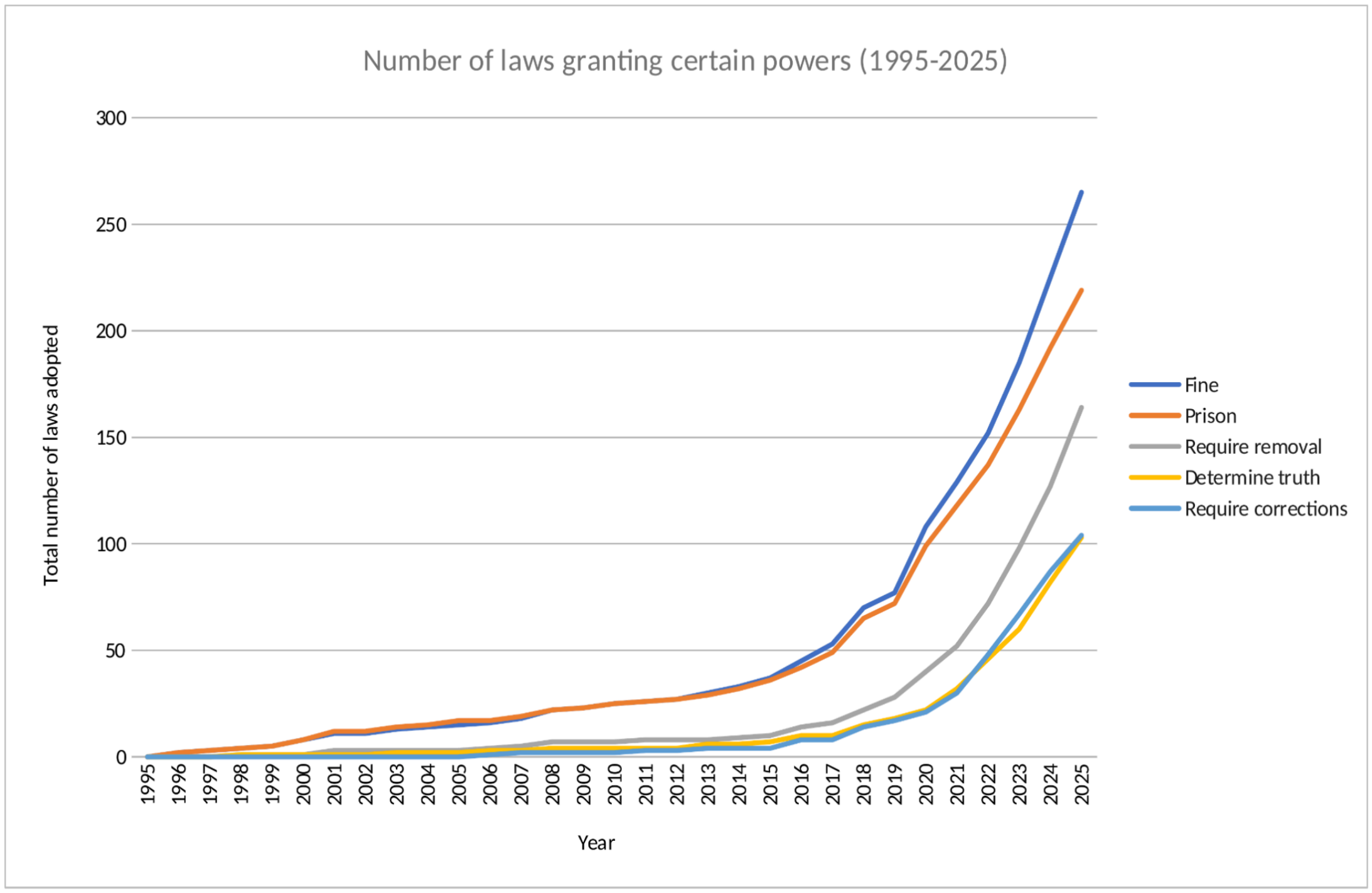

**Figure 8:** Cumulative number of laws granting certain powers between 1995 and 2025

Taken together, the findings indicate that the regulatory response to information disorder relies more heavily on punitive powers than on corrective or remedial measures. Fines and imprisonment were substantially more prevalent than mandatory corrections, content-removal requirements, or truth-determination powers. Although authority to determine whether information is true or false appeared in approximately one-quarter of the instruments, such powers may grant considerable discretion to designated public bodies. As discussed in Section 7.4, these arrangements have faced constitutional scrutiny in several recent cases, particularly where procedural safeguards are limited.

## 8.2. Discretion to Determine Falsity or Harm

The powers granted by an instrument reveal only part of its regulatory design. It is also necessary to identify who decides whether the conditions for using those powers have been met. Similar powers may operate differently depending on whether decisions are made by a minister, an administrative agency, an independent regulator, or a court. This allocation of authority affects institutional independence, transparency, procedural fairness, and access to judicial review.

A preliminary review suggests that instruments authorizing determinations of truth or falsity commonly assign the initial decision to ministers or administrative bodies rather than courts. Courts generally review these decisions on appeal, although in some civil-law jurisdictions they make the initial determination. This may reflect differences in legal and institutional design, but the current free-text coding does not permit that relationship to be tested systematically.

These findings remain exploratory because the dataset records decision-making bodies in free-text fields rather than standardised categories. Future versions could distinguish among ministers, executive agencies, independent regulators, ordinary courts, and specialist tribunals. They could also record whether each body makes the initial decision or reviews an earlier determination. This classification would allow more systematic comparison of how institutional design varies across jurisdictions and regulatory approaches.

## 8.3. Oversight and Judicial Review

Regulatory powers must be considered alongside the safeguards governing their use. Judicial review and appeal can limit administrative discretion, provide remedies for affected parties, and promote procedural fairness. This subsection examines whether instruments that confer truth-determination powers expressly provide for judicial review or appeal.

Among instruments with definitive coding for both variables, 68 of the 97 instruments conferring truth-determination powers (70%) expressly provide for judicial review or appeal. The corresponding proportion is 96 of 240 instruments that do not confer such powers (40%). This difference may indicate that legislatures attach more formal safeguards to powers involving substantial discretion. Alternatively, it may reflect a broader pattern in which more detailed regulatory frameworks specify both wider enforcement powers and more extensive review procedures.

These findings require careful interpretation. An express right of review does not, by itself, establish effective accountability. Its practical value also depends on judicial independence, access to proceedings, the applicable standard of review, available remedies, and compliance with judicial decisions. Conversely, the absence of an express provision does not necessarily mean that review is unavailable under general constitutional or administrative law. Institutional safeguards should therefore be assessed collectively and in their broader legal context, rather than through the presence or absence of any single provision.

## 8.4. Substantive Defences, Exemptions, and Correlations

Procedural review is only one constraint on regulatory authority. Legal defences and exemptions also determine the reach of regulation by specifying when regulatory powers do not apply. Provisions protecting satire, quotation, public-interest reporting, and good-faith mistakes can therefore help reconcile enforcement objectives with freedom of expression.

The dataset records these safeguards as free text in the “limits on application” field rather than as standardised variables. The current analysis is consequently descriptive. Some instruments expressly provide statutory defences or exemptions, whereas others rely on general constitutional or human rights protections. These approaches should not be treated as equivalent. An express statutory defence directly defines the limits of the instrument, while the effect of a general constitutional guarantee depends more heavily on interpretation and enforcement within the wider legal system.

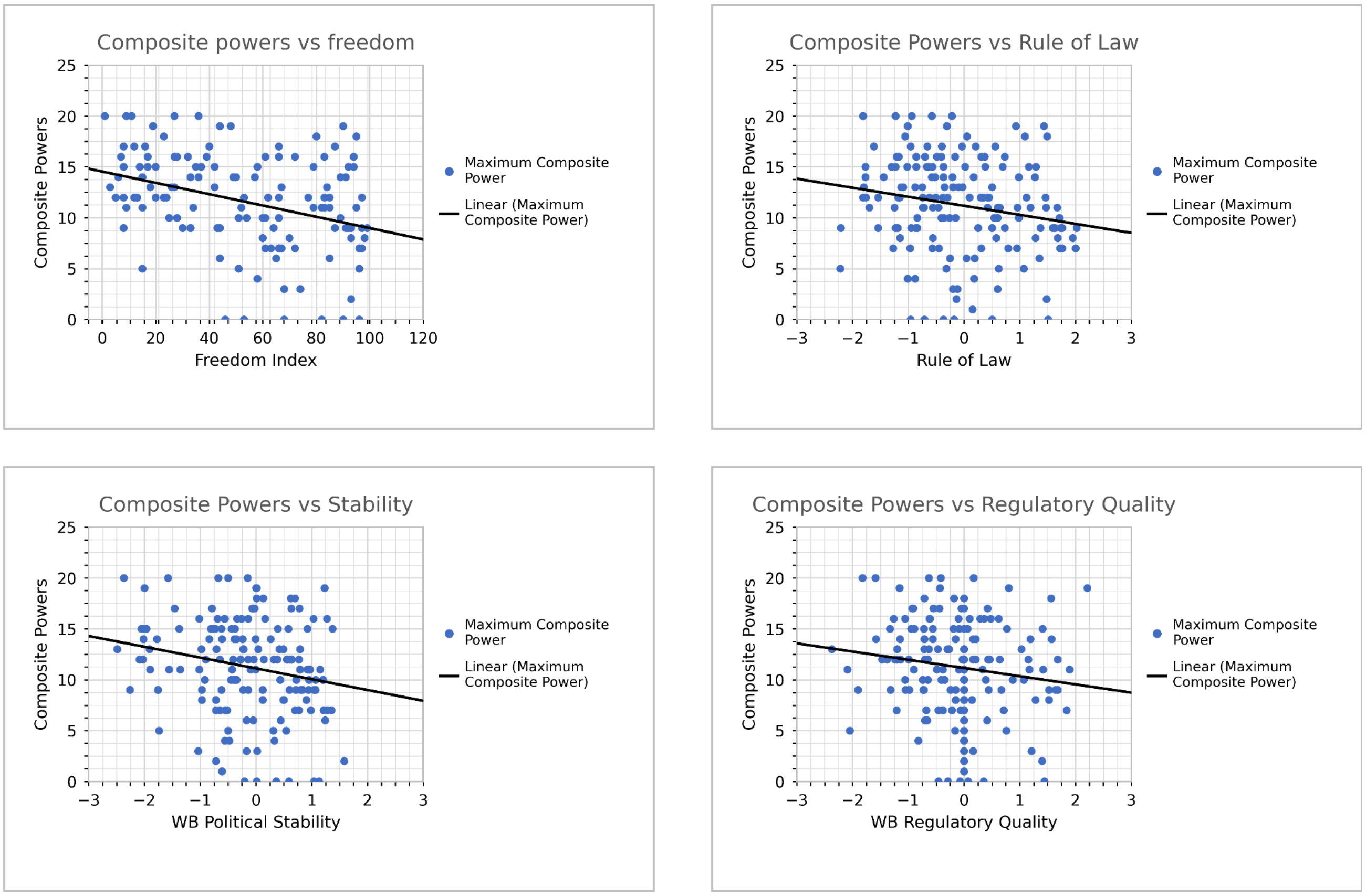

**Figure 9:** Association between the highest Composite Power Index recorded for each jurisdiction and four governance indicators: the Freedom House Global Freedom Score and the World Bank rule of law, political stability, and regulatory quality indicators. Jurisdictions with no Freedom House score recorded are excluded from the first panel.

As Figure 9 shows, all four relationships are negative but weak. Measured at jurisdiction level, the strongest is with the Freedom House Global Freedom Score (r = −0.35, n = 132), followed by political stability (r = −0.21), the rule of law (r = −0.19), and regulatory quality (r = −0.15).

Jurisdictions with lower freedom scores therefore tend to grant somewhat stronger regulatory powers, although the association is modest and is not evidence of a causal relationship.

Overall, governance indicators explain only a small part of the variation in regulatory design across jurisdictions. The scope of regulatory powers and safeguards is more likely to reflect the combined influence of constitutional traditions, legislative choices, and domestic policy priorities.

## 8.5. Transparency, Due Process, and Sunset Provisions

The scope of regulatory authority depends not only on the powers conferred but also on the procedures governing their use. Transparency requirements, due process protections, and sunset provisions can strengthen accountability, reduce arbitrary decision-making, and ensure that regulatory powers remain subject to review. These mechanisms therefore help explain how regulatory frameworks balance enforcement with legal accountability.

Formal due process protections vary considerably across jurisdictions and regulatory approaches. Some instruments expressly require notification, provide for judicial appeal, or administrative review. Others rely on general constitutional or administrative law principles without incorporating such protections into the instrument itself. Judicial review is particularly common where authorities have discretion to determine whether information is true or false. Its presence, however, should be considered alongside the broader regulatory framework and does not, by itself, demonstrate strong institutional accountability.

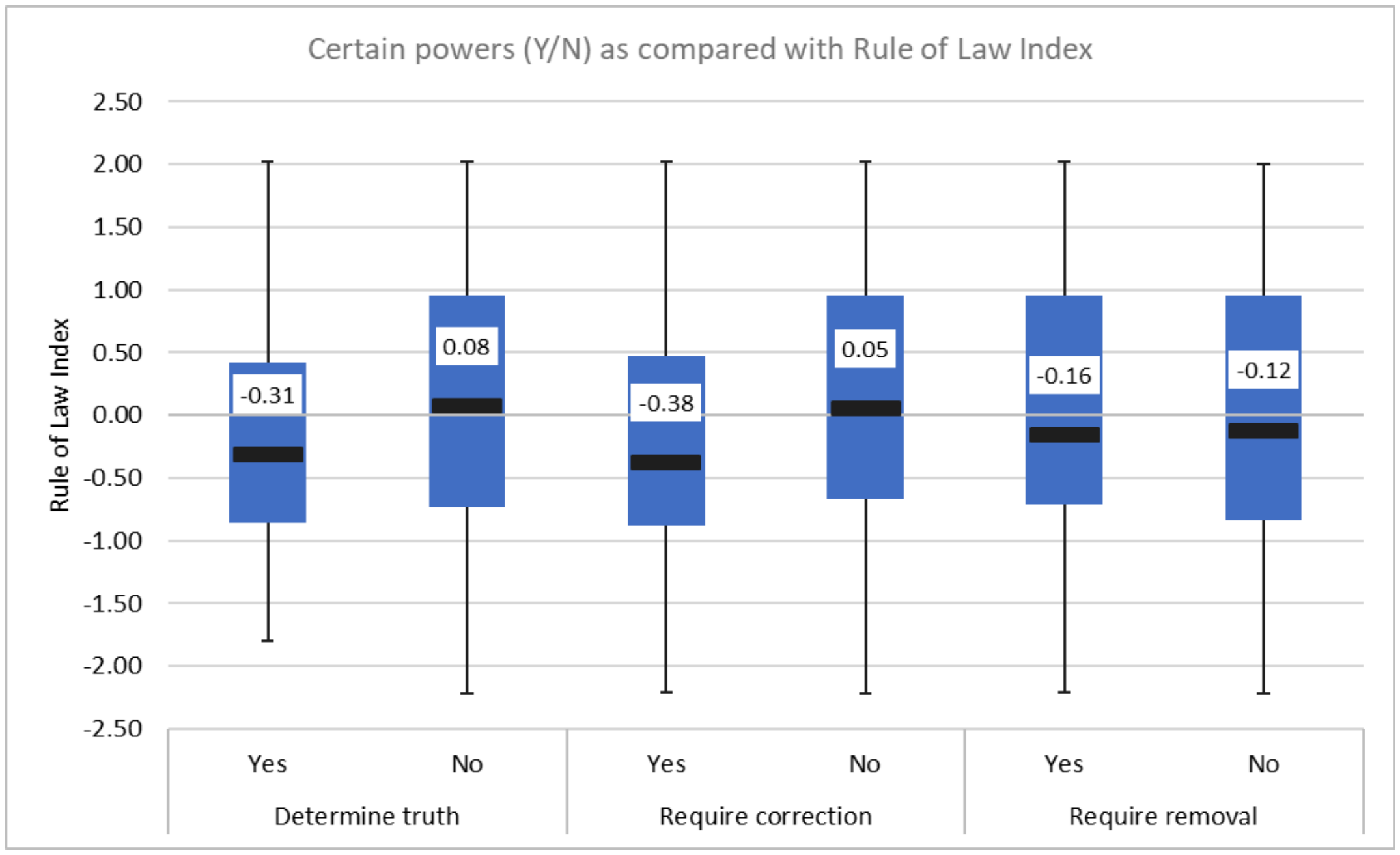

**Figure 10:** Rule of Law Index scores for jurisdictions whose instruments do and do not confer truth-determination, correction, or removal powers (median and interquartile range).

Rule-of-law patterns also differ according to the power examined. Jurisdictions with instruments conferring correction powers have a median Rule of Law score of −0.38, compared with 0.05 for those without such powers. This gap is similar to that observed for truth-determination powers. By contrast, removal powers show almost no difference (−0.16 compared with −0.13). Together with the correlation results reported above, these findings indicate that the relationship between regulatory powers and rule-of-law conditions is not uniform but varies across specific powers.

Transparency obligations are most prominent in platform systems governance. These frameworks commonly require disclosure, reporting, risk assessments, audits, and other compliance measures that support continuing oversight. Accountability is therefore pursued not only through intervention against individual items of content but also through ongoing scrutiny of platform practices.

Sunset clauses and mandatory legislative reviews are uncommon in the dataset, except among measures adopted in response to COVID-19. Many emergency measures included fixed expiry dates and account for a substantial proportion of the 34 instruments classified as repealed or withdrawn. Outside the public health context, statutory limits on the duration of regulatory powers remain rare. Accountability generally depends instead on transparency requirements, review procedures, and continuing institutional oversight.

### 8.6. Sensitivity of the Composite Power Index

The Composite Power Index summarises the breadth and severity of regulatory powers and supports comparisons across jurisdictions. Because its results depend partly on the weights assigned to its components, alternative specifications were tested to assess their robustness.

The index used throughout the report, including in Figures 7 and 9, assigns weights according to the relative severity of each power: coverage = 1, correction = 1, removal = 2, fines = 3, and imprisonment = 5. These weights reflect the assumption that imprisonment constitutes a more severe form of state intervention than administrative measures such as correction or content removal.

As a sensitivity test, an alternative index assigned equal weight to each power by counting the applicable variables. Both specifications produced the same overall regional pattern. The Asia-Pacific and African Groups recorded the highest average scores, while the Western European and Others Group (WEOG) recorded the lowest. The regional comparisons are therefore not materially affected by the weighting scheme.

The Composite Power Index should nevertheless be treated as a comparative measure rather than a precise ranking of regulatory intensity. The consistency of the regional findings across both specifications supports its use for comparison, although alternative weights may still affect the scores of individual jurisdictions.

The regulatory models discussed in Section 7 are becoming central to the discourse on legal intervention against MDM, and Generative AI (hereinafter Gen AI) features prominently within these approaches. The next section examines how Gen AI should be assessed in greater depth. While this assessment partly overlaps with platforms' legal responsibility, it merits separate treatment in shaping future discourse.

## 9. Generative AI and the Next Phase of Regulation

As indicated in Section 5, platform liability has been taken into consideration in regulating intervention against MDM. However, a distinct analysis is required, given the amplification that AI brings to the dissemination of MDM. Social-media platforms were formerly understood primarily as media through which users connect and share ideas. The growing popularity of AI has a greater impact on users in terms of content creation, personalized feeds, and automated data analysis. There is therefore a shifting discourse suggesting that online platforms are no longer neutral, and that AI tools, whether embedded in social media platforms or developed as standalone tools, such as AI chatbots, do not simply generate information but may also be responsible for creating harmful content.

Several cases concerning the liability of internet content providers that were discussed previously, while illustrating the jurisprudence on legal responsibility assigned to online platforms, do not relate specifically to AI and MDM. This database, however, identified various interventions that address the role of Gen AI and how it amplifies MDM. Therefore, while this section discusses the regulatory approach to Gen AI separately, it also considers how current legislation regulates platforms that provide Gen AI as a feature or tool.

## 9.1. Deepfakes and Impersonation

By definition, a deepfake is created using AI, as it relies on deep learning technology to generate or manipulate images, audio or video. However, long before AI tools became popular, false information, such as fake images or impersonation, was already being produced through tools such as video or photo editing.

AI tools make deepfakes easier to produce at scale, increasing the volume of synthetic images, audio and video. This form of technological abuse is evident in cases that harm the reputation and privacy of victims, since no consent is given for generating such online content.

Denmark proposed amendments to its Copyright Act in 2025 to protect against realistic digitally generated imitations of a person's physical characteristics, including appearance and voice. It would be the first law against deepfakes to grant individuals exclusive control over their physical features and voice as intellectual property. A year earlier, France introduced a bill on AI-generated images, although it focuses on the mandatory labelling of AI-generated or modified images on social networks.

## 9.2. Intimate and Sexual Imagery

Regarding the creation of child sexual abuse material in particular (see the discussion in Section 4), a growing number of laws are targeting these offences against children, although most focus on the criminal harm regulatory model. This emphasis is understandable given the rising number of cases,[51] some of which are extreme, sophisticated and evidently cause severe harm to the victims.[52]

Australia has proposed such an approach. Its Criminal Code Amendment Bill includes changes to provisions concerning technology used to generate child abuse material.[53] Interestingly, while the Bill does not focus on regulating platforms, its provisions seek to criminalize individuals who download, access, supply, or train AI technology for child abuse material.

---

[51] Internet Watch Foundation. (2026). *Harm Without Limits: AI Child Sexual Abuse Material Through the Eyes of Our Analysts,* Available at https://www.iwf.org.uk/media/hl1nvdti/iwf-ai-csam-report-2026.pdf.
[52] Karasavva, V. (2025). The Frequency, Nature, Impact, and Coping Strategies of Nonconsensual Intimate Image Dissemination Victimization: A Scoping Review. *Trauma, Violence, & Abuse*, 0(0). 11, 1-16.
[53] Criminal Code Amendment (Using Technology to Generate Child Abuse Material) Bill 2025 (Cth).

Meanwhile, taking a less binding approach, Andorra passed a Code of Ethics for Artificial Intelligence.[54] Focusing on trustworthy AI and a human rights framework, the Code regulates AI stakeholders, developers, deployers, and users, as well as entities involved in public procurement.

## 9.3. Electoral Synthetic Media

In several recent incidents involving electoral integrity, the harm to the electoral process is no longer confined to foreign interference that causes polarization and influences voter behaviour. The discussion has shifted to the growing dissemination of synthetic content before and during elections.

During Argentina's 2023 presidential election, both presidential candidates, Sergio Massa and Javier Milei, used AI tools to create deepfakes for their campaigns[55]. Meanwhile, before Singapore's 2025 General Election, a video of the late Prime Minister Lee Kuan Yew circulated online[56]. These examples illustrate the growing use of synthetic content in electoral campaigning and political messaging.

To date, Argentina has no legislation specifically regulating elections and MDM, as the country focuses instead on two policies. Singapore, by contrast, has passed six different laws. One of these focuses on the distribution of MDM relating to elections: it bans manipulated online election advertising that contains realistic but false representations of candidates, as well as making of false or misleading declarations.[57]

## 9.4. Content Provenance and Watermarking

As discussed in the previous section, a growing body of legislation classifying the authenticity of online content is now recognized across several jurisdictions. The approach of disclosing who created the content may not directly reduce the dissemination of MDM, given that it was implemented largely to educate the public about the existence of synthetic content online.

However, legislation on content provenance and watermarking becomes significant when it focuses on creators and platforms as the responsible actors, particularly in disclosing their contribution to producing such content. Consequently, when harm affects protected interests or victims, the legal system can (at least in theory) more easily assess the legal responsibility arising from the circulation of synthetic content.

There are a few examples of legislation that focus on platforms, and even on AI developers. France's 2024 bill, discussed earlier in this section, is one of them; it focuses on the mandatory labelling of AI-generated or modified images on social networks and on platforms' detection obligations. China's 2025 regulation, titled "Measures for labelling of AI-Generated Synthetic Content" (see Section 6.5), is a similar example.

## 9.5. AI-generated Text and Chatbot Outputs

---

[54] Government of Andorra, *Andorran Code of Ethics for Artificial Intelligence*, 2024, Available at https://andorra-digital.com/images/pdf/Codi%20etic%20english.pdf

[55] Łabuz M, Nehring C. On the way to deep fake democracy? Deep fakes in election campaigns in 2023. *European Political Science*. 2024;23(4): 454-473. 461. doi:10.1057/s41304-024-00482-9.

[56] Channel NewsAsia, *Rise of AI and Deepfakes in Leadup to GE 2025*, CNA, 2025, https://www.channelnewsasia.com/interactive/ge2025-deepfake/.

[57] Elections (Integrity of Online Advertising) (Amendment) Act 2024 (No. 34 of 2024), Section 61MA (Singapore).

The previous subsections discussed emerging issues attached to AI developers, although these stem from the broader discussion on online platforms. Given the newness of AI chatbot platforms such as ChatGPT and Claude, the regulation of these platforms requires separate discussion, especially because their tasks and functions differ from those of online platforms like social media.

The term "AI hallucination" refers to instances where AI systems generate false or misleading information and present it as fact[58]. The emergence of search engine platforms that embed AI features is currently widespread. However, several of the regulations found in the database are fairly general, addressing platforms' responsibility for online synthetic content. In Iran, a 2025 Bill on Combating the Spread of False News Content in Cyberspace proposed criminal penalties for knowingly disseminating false online content, alongside platform duties concerning fact-checking, warning labels, corrections and removal. After parliament approved expedited consideration, the government withdrew the bill on 30 July 2025.

Similarly to China's 2025 regulation, which requires the labelling of synthetic content, Romania introduced a Bill that regulates the malicious use of technology and limits the deepfake phenomenon through labelling and prohibitions.

Various cases involving AI hallucination have occurred across different jurisdictions,[59] and the approaches taken by legal systems vary. Many of these cases involve fabricated sources submitted in court proceedings. However, much of this misuse of AI tools has been sanctioned with either a monetary fine, a warning, or a bar referral. This shows that regulation of AI chatbot output is genuinely needed, but that the law still has to catch up on how to formulate adequate sanctions, particularly in professional contexts such as legal practice.

## 9.6. Responsibility of Foundation-model Providers

The way AI developers build their platforms and train their models as part of the platform's function refers, in this context, to how AI platforms ensure the responsibility and transparency of the tools they develop. This includes risk assessment and mitigation, such as content moderation and filters, and transparency about how the training data was used, including efforts to maintain fairness and reduce bias in the information generated.

Moreover, the tasks assigned to these AI tools should ensure that the tools incorporate security safeguards, are not open to abuse, and provide a mechanism for users to report problems, as well as redress for harm as part of the AI developer's accountability, and compliance with any national and international legislative frameworks related to data privacy or copyright.

AI developers themselves have issued internal guidelines and policies to ensure responsibility and transparency in the models they provide. Anthropic has published Claude's Constitution,[60] which relies on supervised learning and reinforcement learning from AI feedback, while OpenAI provides a Moderation API that developers can use to classify potentially harmful inputs or outputs.[61]

---

[58] *See* Sun, Y., Sheng, D., Zhou, Z. *et al.* AI hallucination: towards a comprehensive classification of distorted information in artificial intelligence-generated content. *Humanit Soc Sci Commun.* 11, 1278 (2024). https://doi.org/10.1057/s41599-024-03811-x.

[59] See Matthew Dahl, Varun Magesh, Mirac Suzgun, Daniel E Ho, Large Legal Fictions: Profiling Legal Hallucinations in Large Language Models, *Journal of Legal Analysis*, Volume 16, Issue 1, 2024, Pages 64–93, https://doi.org/10.1093/jla/laae003; *See also database of AI hallucination cases at* Damien Charlotin, *AI Hallucination Cases* (n.d.), available at https://www.damiencharlotin.com/hallucinations/.

[60] Anthropic, *Claude's Constitution* (May 9, 2023), available at https://www.anthropic.com/news/claudes-constitution.

However, with regard to compliance with international human rights standards, these companies' policies still face challenges in ensuring alignment. A 2024 analysis by the Future of Free Speech argues that leading generative-AI content policies often fall short of the legality, legitimacy and necessity standards used in Article 19 ICCPR analysis.[62]

## 9.7. Authenticity of Human-generated Content

The previous section outlined that provenance regulation is one of the regulatory models implemented across various jurisdictions. Governments impose these regulations through both hard and soft law. Platforms, however, also need to introduce such regulation to ensure accountability within their services. Given the emerging nature of this regulatory model, various pieces of legislation in the database combine several models within a single instrument, such as criminal sanctions, risk assessment, and provenance regulation.

More importantly, distinguishing humans from machines appears to be one of the common approaches adopted by AI platforms, particularly through labelling whether content was created by a human and uploaded to the platform or generated entirely by the system and distributed there.

This approach aligns with recent state regulatory efforts, as seen in Denmark's copyright amendment. As discussed earlier, such efforts aim to prevent impersonation and deception. They also help maintain consumer trust in the online content that people consume, and, more broadly, protect the integrity of the information ecosystem by safeguarding the originality of content and the authenticity of the actors involved in the digital information pipeline.

A few examples illustrate this. SynthID is Google DeepMind's technology, which embeds invisible watermarks. Another is the Coalition for Content Provenance and Authenticity (C2PA), which develops an open technical standard for recording and verifying provenance information. It is worth noting, however, that what these tools actually verify is the source and integrity of a file, which device or tool produced it, and that it has not been modified since, rather than the deeper question of whether it was genuinely made by a human without AI assistance and is, in a meaningful sense, true.

## 9.8. Likely Direction of the Proposal Pipeline

Given that MDM remains a global, cross-jurisdictional threat, the question of how to regulate it effectively, whether through a strict or a lenient approach, will likely continue to arise in efforts to combat its dissemination. The six regulatory models currently implemented (discussed in Section 7) stem partly from classical, existing legislative systems, such as policy instruments and criminal codes, and partly from new and emerging trends, such as provenance regulation and risk assessment.

However, given the potential for new and varied developments in AI as a technology, the regulatory approaches identified in this study may well change in the future. Criminal law remains the most common legislative model, but in light of concerns about its effectiveness and its potential for abuse and adverse effects on human rights, alternative approaches are being pursued, as suggested by recent examples such as India's Fact-Check Unit and Australia's Disinformation Bill. Criminal responsibility of platforms and developers remains a subject of debate, raising the tension between potential benefits of AI and minimizing/mitigating its risks.

---

[61] OpenAI, *Moderation* (n.d.), https://developers.openai.com/api/docs/guides/moderation.

[62] Jordi Calvet-Bademunt & Jacob Mchangama, *Freedom of Expression in Generative AI: A Snapshot of Content Policies* (The Future of Free Speech, 2024), https://futurefreespeech.org/wp-content/uploads/2023/12/FFS_AI-Policies_Formatting.pdf.

It will be interesting to see whether future measures are enacted as separate, dedicated instruments of hard law by governments, rather than as internal policies or guidelines issued by platforms, or as regulations combining different legislative models. Against this backdrop, the next section will discuss the implementation of existing legislative frameworks on MDM and how these legislative models might develop in the future.

# 10. Law in Action: Implementation, Contestation, and Effectiveness

The preceding analysis has focused primarily on the formal regulatory framework: the measures governments have adopted, their design, and the powers and safeguards they establish. The dataset therefore provides extensive evidence on regulatory design but less systematic evidence on implementation and outcomes.

This section begins to address that gap through selected enforcement records, judicial decisions, and independent analyses. These sources reveal emerging patterns in enforcement, legal challenges, institutional capacity, and regulatory effects. The analysis remains necessarily selective because comparable implementation data are unavailable for many jurisdictions and instruments.

## 10.1. Selected Case Studies in Implementation

Five instruments spanning several of the regulatory models identified in Section 7 illustrate the range of implementation experiences documented in available sources. Singapore's Protection from Online Falsehoods and Manipulation Act (POFMA), adopted in 2019, is a prominent example of the truth-policing model. As of 1 July 2026, official figures record 92 cases, which together resulted in 146 POFMA Correction Directions[63]. Most required correction notices rather than content removal, fines, or imprisonment, reflecting the Act's emphasis on correction as its primary remedy. However, repeated action against a small number of respondents has raised concerns about escalation and selective application.

While it remains controversial due to the tension between protecting free speech and preventing the harmful effects of MDM, the criminal law approach is still the most common means of preventing the large-scale distribution of MDM. However, as this study suggests, various regulatory frameworks are gradually emerging (see Table 3), highlighting a combined approach to combating MDM. China's Criminal Law has criminalised certain forms of false information since before the COVID-19 pandemic. Criminal Law Amendment (IX), adopted in 2015, added a provision criminalizing the fabrication and dissemination of false information concerning emergencies, epidemics, disasters or police situations where this seriously disrupts public order. The provision, now contained in Article 291a of the Criminal Law, became particularly relevant during the COVID-19 pandemic. However, a decade later, the 2025 Regulation predominantly introduced provenance rules, referring to other existing laws for punishment (see discussion in section 6.5).

---

[63] Government of Singapore. (n.d.). *Tabulation of POFMA cases and actions*, POFMA Office, available at https://isomer-user-content.by.gov.sg/501/80bbf192-e278-4ab9-93cc-8da479af0819/tabulation_of_pofma_cases_and_actions.pdf.

Regarding the implementation of the Criminal Law amendment, documented enforcement extends broadly across cyber-related offences rather than being confined to false-information offences. According to the Supreme People's Procuratorate, 323,000 people were prosecuted in 2023 for crimes committed using the internet, encompassing conduct ranging from online rumours, defamation, and doxxing to telecommunications fraud, assisting information-network crimes, and online gambling.

Meanwhile, with respect to the Measures for labelling of AI-Generated Synthetic Content, In November 2025, the CAC announced enforcement action against a batch of mobile applications for non-compliance with AI-generated content-labelling requirements, although it did not disclose the number of applications involved. In April 2026, the CAC publicly identified three platforms—Jianying, Maoxiang, and Jimeng AI—as having failed to comply with AI-generated content-labelling requirements and imposed regulatory measures including warnings and rectification orders[64]. Broader CAC campaigns have reported substantially larger enforcement figures, but these aggregate statistics encompass multiple forms of AI-related non-compliance and cannot be attributed specifically to violations of the Measures.

The EU Digital Services Act (DSA) provides the clearest implementation record among the newer platform-governance models. The European Commission opened formal proceedings against X in December 2023 and issued its first non-compliance decision in December 2025.[65] The decision imposed a €120 million fine for deceptive verification practices and deficiencies in the platform's advertising repository and researcher-access arrangements. The Commission also secured binding commitments from TikTok on related transparency concerns without imposing a fine.[66] Other investigations, including proceedings concerning the protection of minors on adult-content platforms, remained ongoing at the time of writing. These developments show that DSA enforcement has expanded gradually since the Act entered into force.

South Korea's September 2024 amendments to the Act on Special Cases Concerning the Punishment of Sexual Crimes combine elements of victim-centred remedies and criminal harm prevention. The amendments extended criminal liability to the possession and viewing of sexually explicit deepfakes, rather than limiting it to their production or distribution. The subsequent enforcement record illustrates the increasing use of criminal law to address emerging forms of technology-facilitated sexual harm, although independent analysis indicates that sentences have often remained well below the statutory maximum.

| Regulatory model | Instrument | Documented implementation outcome |
|---|---|---|

[64] Cyberspace Administration of China (CAC), 网信部门依法查处“剪映”App等生成合成内容标识违法问题网站平台 [Cyberspace Authorities Investigate and Penalize Jianying App and Other Online Platforms for Violations of AI-Generated and Synthetic Content Labelling Requirements], April 28, 2026, available at https://www.cac.gov.cn/2026-04/28/c_1779119736411711.htm.

[65] Anna Pingen, *EU Fines X €120 Million in First DSA Non-Compliance Decision* (2025), Eu Crim, available at https://eucrim.eu/news/eu-fines-x-120-million-in-first-dsa-non-compliance-decision.

[66] International Association of Privacy Professionals, *European Commission Fines X €120M Euros for DSA Violations* (2025), IAPP, available at https://iapp.org/news/a/european-commission-fines-x-120m-euros-for-dsa-violations.

| | | |
|---|---|---|
| Truth-policing | Singapore, POFMA (2019) | 92 cases and 146 correction directions as of 1 July 2026. Most required correction rather than removal, fines, or imprisonment. |
| Criminal harm prevention | China, Criminal Law of the People's Republic of China (Amendment IX, 2015) | In 2023, there were 323,000 prosecutions, although this figure reflects cyberspace offences broadly, not the criminalization of false information alone. |
| Provenance regulation / criminal harm prevention | China, Measures for labelling of AI-Generated Synthetic Content (2025) | Three platforms were publicly identified in April 2026 for violations expressly involving the Measures. Broader AI-enforcement statistics are available but are not disaggregated by violations of the Measures. |
| Platform systems governance | EU, Digital Services Act | First non-compliance decision issued in December 2025, imposing a €120 million fine on X almost two years after formal proceedings began. The Commission also secured binding commitments from TikTok, while further investigations continued in 2025–2026. |
| Victim-centred remedies / criminal harm prevention | South Korea, deepfake-sex-crime amendments (2024) | Reports and apprehensions rose sharply in 2024 amid increased enforcement activity and the establishment of a dedicated police task force. Analysis of court decisions from 2020–2024 nevertheless identified continued leniency in sentencing relative to statutory maximums. |

**Table 5:** Selected implementation case studies by regulatory model.

## 10.2. Enforcement Patterns

The selected cases reveal three broad implementation patterns that are not apparent from formal regulatory powers alone.

First, truth-policing measures such as POFMA can generate frequent, lower-level administrative actions. Correction directions are issued regularly, while financial or criminal penalties are generally associated with repeated or more serious non-compliance. India and Mongolia show, however, that similar truth-policing powers may also face constitutional challenges before they become fully operational.

Second, platform systems governance involves fewer but potentially more complex investigations. Under the DSA, enforcement may require regulators to obtain extensive information and assess platform-wide practices and systems. Almost two years elapsed between the Commission's opening of formal proceedings against X in December 2023 and its first non-compliance decision in December 2025. During that period, the Commission issued further information requests, required the preservation of internal documents, and sought access to X's APIs. This illustrates the evidential and technical demands of systems-level platform oversight, although significant parts of the investigation remained ongoing after the first decision.

Third, measures combining criminal harm prevention with victim-centred remedies depend heavily on enforcement capacity. South Korea recorded approximately 156 reports of deepfake sexual offences in 2021 and 812 reports in the first nine months of 2024, during which 387 suspects were apprehended.[67] By comparison, approximately 16 arrests were recorded among 793 reported cases during the preceding three years. The sharp increase in 2024 coincided with greater public attention and the establishment of a dedicated police task force. Because the legislative amendments were adopted only in September 2024, these figures should not be interpreted as evidence of their effects. Instead, they provide a baseline against which subsequent implementation can be assessed and suggest that reporting and enforcement capacity may previously have constrained the law's application.

[67] Georgia Smith & Joseph Brake, *South Korea Confronts a Deepfake Crisis* (Nov. 19, 2024), available at https://eastasiaforum.org/2024/11/19/south-korea-confronts-a-deepfake-crisis/.

## 10.3. Constitutional and Judicial Challenges

The implementation record supports the earlier finding that truth-policing measures are particularly vulnerable to constitutional challenge. India's Fact-Check Unit was invalidated before becoming operational, while Mongolia's existing criminal false-information provision was struck down in 2025. In both cases, the courts expressed concerns about the breadth or clarity of the provisions and the safeguards governing state authority to determine whether online content was true or false. These decisions align with the broader association identified in the dataset between truth-determination powers and weaker rule-of-law conditions, although they do not establish a causal relationship.

Platform systems governance has produced a different pattern. None of the instruments examined in this category has been invalidated by a court. Instead, X has challenged the European Commission's December 2025 non-compliance decision on procedural and substantive grounds. The case will provide an early indication of how EU courts assess transparency and systems-level obligations imposed directly on platforms, as distinct from state decisions about the truth or falsity of individual claims.

This contrast reflects an important difference in regulatory design. Truth-policing measures place public authorities in the position of judging contested information, creating direct tensions with freedom of expression and procedural fairness. Platform systems governance generally regulates the processes, risks, and responsibilities of intermediaries without giving governments direct authority to determine the truth of individual claims. Whether this distinction continues to influence judicial outcomes will become clearer as more platform-related cases reach the courts.

## 10.4. Evidence of Impact and Unintended Consequences

The case studies in Sections 10.1–10.3 establish whether selected instruments have been implemented or challenged. A more difficult question is whether they have achieved their intended objectives or produced unintended effects. The available evidence is limited and illustrative, but three patterns are apparent.

First, correction-based remedies such as Singapore's POFMA leave the original content accessible while attaching an official correction. Their effectiveness therefore depends partly on whether users encounter, read, and accept the correction—outcomes that the dataset cannot measure. The repeated use of POFMA against a small number of respondents also shows that enforcement may become concentrated among particular individuals or organizations. The available evidence, however, does not establish whether this pattern reflects selective enforcement, repeated non-compliance, or both.

Second, South Korea's response to deepfake-related sexual offences illustrates the difference between legal authority, enforcement activity, and judicial outcomes. Reports and apprehensions rose sharply during 2024 amid greater public attention and the establishment of a dedicated police task force. Because the legislative amendments were adopted only in September 2024, figures covering the first nine months of that year cannot demonstrate the effects of removing the intent requirement. The increase may instead reflect a combination of greater reporting, stronger enforcement, institutional prioritization, and changes in offending. At the same time, analysis of court decisions from 2020–2024 indicates that sentences often remained well below the statutory maximum. Stronger investigative and charging powers therefore do not necessarily produce more severe sentences—a distinction that the Composite Power Index cannot capture.

Third, changes in the wider digital information environment may affect every regulatory model. Meta ended its US third-party fact-checking programme in 2025 and replaced it with Community Notes. Such changes may alter how false or manipulated content is challenged, although their effects remain to be established. Reduced private investment in verification and moderation could place greater pressure on public correction and removal mechanisms while testing the capacity of platform systems governance.

### 10.5. Regulatory Capacity and Resources

Section 7.3 identified almost no relationship between the Composite Power Index and government effectiveness ($r \approx -0.03$) and only a weak negative relationship with the rule of law ($r \approx -0.26$). Together with the evidence in Sections 10.1–10.3, these results suggest that formal legal powers do not by themselves indicate enforcement capacity. Although the Asia-Pacific and African regional groups recorded the highest average Composite Power Index scores, the most detailed case studies come from jurisdictions with extensive public reporting, particularly Singapore's POFMA administration and the European Commission's DSA enforcement system. This creates an important limitation. Observable enforcement may be concentrated in jurisdictions with stronger administrative and reporting capacity rather than accurately representing implementation across the dataset's full geographic coverage.

Where capacity constraints are visible, they appear to affect the scale, speed, and consistency of enforcement. In South Korea, the rise in apprehensions during 2024 coincided with the establishment of a dedicated police task force, suggesting that institutional prioritization and investigative resources can influence enforcement. However, because legislative changes were adopted during the same period, the separate effects of legal reform and increased capacity cannot yet be determined. Similarly, the almost two years between the opening of proceedings against X and the first DSA non-compliance decision illustrate the time, expertise, and evidence required to investigate systemic platform failures. Such demands may limit the number of investigations that regulators can pursue simultaneously.

The dataset does not currently record regulatory capacity or resources as structured variables. Its free-text "Accountability/oversight" field identifies institutional arrangements but does not consistently capture staffing, funding, technical expertise, or caseloads. A structured capacity indicator should therefore be developed for the mid-2027 update. Relevant measures could include the existence and establishment date of dedicated enforcement units, staffing and funding levels, caseloads, and published enforcement outcomes. These measures may provide a better indication of implementation capacity than formal legal powers alone.

## 11. Conclusions and Future Directions

This report has examined the growth, geographic distribution, substantive focus, regulatory design, and early implementation of national responses to misinformation and disinformation between 1995 and 2025. This concluding section draws together the principal findings. It identifies areas of convergence and continuing variation, considers the political and constitutional durability of different regulatory models, highlights gaps in regulation and the available evidence, and outlines priorities for updating the dataset in mid-2027 and mid-2028.

## 11.1. Convergence and Persistent Divergence

The dataset reveals two forms of convergence. The first concerns adoption. Since 2016, the gap between earlier-adopting regions, particularly Africa and the Asia-Pacific, and later-adopting regions, notably WEOG, has narrowed substantially. Regulation of misinformation and disinformation has consequently become a widespread global policy response rather than a feature of only a few regions (Section 3.3).

The second form concerns regulatory design. As the six-model typology in Section 7.1 demonstrates, recent hard-law instruments increasingly combine elements of several models. A broadly similar set of measures—including platform transparency obligations, remedies for affected individuals, and content-provenance requirements—now appears across jurisdictions with different legal traditions. Examples range from the European Union's Digital Services Act and South Korea's deepfake legislation to measures introduced across Africa and the Asia-Pacific since 2024 (Section 7.2).

This convergence has not eliminated important differences in the breadth of regulatory powers or the institutions responsible for exercising them. The Composite Power Index shows that instruments in the Asia-Pacific and African Groups generally confer broader powers than WEOG instruments addressing comparable harms. Legal traditions also shape institutional design. Common-law jurisdictions more frequently assign content assessment and platform oversight to administrative bodies, whereas civil-law jurisdictions more often involve courts. Eastern European jurisdictions also rely more heavily than other regions on broadly framed criminal offences concerning false information (Section 7.3). Governments are therefore adopting an increasingly similar regulatory toolkit, but substantial differences remain in the strength of intervention and the allocation of enforcement authority.

## 11.2. Political and Constitutional Sustainability

The six regulatory models face different constraints on their long-term durability. Truth-policing measures, which authorise public bodies to determine whether information is true or false, appear particularly vulnerable to constitutional challenge. The invalidation of India's Fact-Check Unit and Mongolia's criminal false-information provision illustrates this risk. At the aggregate level, the dataset also associates truth-determination powers with somewhat weaker rule-of-law conditions, although this relationship does not establish causation.

Where truth-policing measures survive judicial review, their durability will depend not only on their formal legality but also on how their powers are used. Singapore's POFMA has remained in force, but its repeated application to a relatively small group of respondents warrants continued examination. Patterns of enforcement, the availability of review, and the treatment of repeat respondents will be important to assessing its longer-term legitimacy.

Platform systems governance faces different pressures. No platform-governance instrument coded in the dataset had been judicially invalidated by the end of the study period. Nevertheless, Australia's withdrawal of its Combating Misinformation and Disinformation Bill after sustained political opposition shows that such measures require continuing political support. The lengthy investigation preceding the European Union's first Digital Services Act non-compliance decision also demonstrates the substantial administrative capacity required for implementation.

Victim-centred and provenance-based measures are newer and less common in the dataset, making firm conclusions about their durability premature. South Korea's experience nevertheless shows that governments may respond to enforcement shortfalls by expanding liability and investing in specialised enforcement capacity. Whether these measures produce sustained improvements while preserving procedural safeguards requires further study.

## 11.3. Gaps in Existing Regulation

The analysis identifies several gaps in the current regulatory landscape. First, synthetic text remains more difficult to identify and label consistently than synthetic images, audio, and video. Provenance requirements continue to focus primarily on visual and audiovisual content. AI-generated text and chatbot outputs therefore remain comparatively less regulated, even as generative AI systems produce them at increasing scale.

Second, the national focus of the dataset does not fully capture the cross-border character of information disorder. Digital platforms operate across many jurisdictions, while most regulatory powers remain territorially limited. Transatlantic disputes surrounding the European Union's enforcement action against X illustrate the resulting tensions. Future research should examine how national and regional regulatory systems interact, overlap, and potentially conflict.

Third, several aspects of institutional design are not yet captured through variables that allow consistent comparison. Information about the bodies authorised to determine truth, available defences and exemptions, and oversight and accountability arrangements is recorded mainly in free-text fields. Converting these fields into structured variables would enable closer analysis of how institutional choices affect enforcement and legal outcomes.

The most significant gap concerns effectiveness. Available data demonstrate that enforcement occurred but not whether it reduced harm, improved information quality, or produced unintended effects. A related gap concerns regulatory capacity. The difference between the powers formally granted to governments and their practical ability to exercise those powers is currently visible mainly through individual cases. Systematic information about staffing, funding, expertise, caseloads, and enforcement outcomes is needed to explain why similarly designed instruments may operate differently across jurisdictions.

## 11.4. Generative AI and Information Authenticity

Generative AI is likely to shape the next phase of regulatory development. The dataset already shows increasing attention to AI-generated content, but implementation raises unresolved questions. As generative systems produce increasingly convincing text, audio, images, video, and combinations of these formats, existing labelling and provenance requirements may need to extend beyond their current emphasis on visual content. Regulatory frameworks must also allocate responsibility more clearly among platforms, developers of general-purpose AI systems, content creators, and users.

Legislation alone cannot resolve these questions. The practical strengths and limitations of emerging approaches will become clearer only as new measures are implemented, enforced, and reviewed. The proposal pipeline already indicates growing attention to platform governance, provenance, and AI transparency.

Future updates will track these developments and assess how regulatory approaches evolve alongside generative technologies and their associated risks. AI developers may adopt internal policies, while states enact binding legislation. Future updates will examine how these companies comply with laws across jurisdictions.

## 11.5. An Agenda for Continued Data Collection and Evaluation

The findings establish six priorities for the next phase of the project. First, the dataset will continue to be maintained on an annual cycle, with substantive updates planned for mid-2027 and mid-2028. Regular updates will allow regulatory developments to be recorded as they occur rather than reconstructed retrospectively.

Second, the free-text fields covering informational harm and institutional design, discussed in Sections 4.5, 8.2 and 8.4, will be converted into structured variables. Future versions will systematically identify the protected interests within the legislative provisions, responsible decision-making body, available defences and exemptions, and mechanisms for oversight and accountability. This will support more consistent comparison across jurisdictions.

Third, the dataset will begin recording implementation and enforcement outcomes where reliable public information is available. These variables will cover correction directions and similar administrative actions, fines and other penalties, prosecutions and their outcomes, and judicial decisions concerning the validity or application of regulatory measures. This “law in action” component will complement the dataset’s existing focus on formal legal design.

Fourth, future analysis will examine several explanatory factors simultaneously. These will include political freedom, government effectiveness, legal tradition, regional grouping, and, where reliable measures can be developed, regulatory capacity. This approach should help distinguish the factors associated with regulatory design from those that influence implementation and enforcement.

Fifth, the search methodology described in Section 2.5 will be expanded to reduce reliance on English-language sources. Better use of local-language materials should improve coverage and reduce the risk that jurisdictions with limited English-language documentation are under-represented.

Finally, the project will continue to monitor the proposal pipeline identified in Section 3.6 as an early indicator of regulatory change. Particular attention will be given to proposals concerning AI-generated text and audio, for which the regulatory gaps identified in this report are likely to become increasingly important.

## 11.6. Beyond Falsehoods

The regulatory landscape is therefore moving beyond a narrow concern with whether information is true or false. Increasingly, governments are regulating the harms, actors and systems through which information is created, amplified and consumed. Whether these newer approaches prove more effective, proportionate and sustainable remains uncertain. Tracking that evolution — in law and in practice — will be the central task of future editions of this report.